\documentclass[a4paper,fleqn,usenatbib]{mnras}

\usepackage{amssymb}
\usepackage{amsmath}
\usepackage{graphicx}
\usepackage{natbib}
\usepackage{hyperref}
\usepackage{pdflscape}
\usepackage{longtable}
\usepackage{multicol}
\usepackage{multirow}
\usepackage{url}
\usepackage{graphicx}
\usepackage{color}
\usepackage{txfonts}

\DeclareRobustCommand{\ION}[2]{%
\relax\ifmmode
\ifx\testbx\f@series
{\mathbf{#1\,\mathsc{#2}}}\else
{\mathrm{#1\,\mathsc{#2}}}\fi
\else\textup{#1\,{\mdseries\textsc{#2}}}%
\fi}
\newcommand{\lam}{$\lambda$}

\newcommand{\nii}{[\ION{N}{ii}]}

\newcommand{\oii}{[\ION{O}{ii}]}
\newcommand{\oiii}{[\ION{O}{iii}]}
\newcommand{\sii}{[\ION{S}{ii}]}

\newcommand{\Ha}{$\rm{H}\alpha$}
\newcommand{\Hb}{$\rm{H}\beta$}

\newcommand{\temp}{T$_{\rm e}$}

\title[MZ relation revisited]{The Mass-Metallicity Relation revisited with CALIFA} 

\author[S.F.S\'anchez et al.]{S.F. S\'anchez$^{1}$, J.K. Barrera-Ballesteros$^{2}$, 
L. S\'anchez-Menguiano$^{3,4}$, C.~J.~Walcher$^{5}$,
\newauthor 
R. A. Marino$^{6}$, L. Galbany$^7$, J.~Bland-Hawthorn$^8$, M. Cano-D\'\i az$^{1}$, R. Garc\'\i a-Benito$^{3}$,
\newauthor
C. L\'opez-Cob\'a$^{1}$, S. Zibetti$^9$, J.M. Vilchez$^{3}$, J. Igl\'esias-P\'aramo$^{3}$, C. Kehrig$^3$,
\newauthor
A.~R.~L\'opez S\'anchez$^{10,11}$, S. Duarte Puertas$^3$, B. Ziegler$^{12}$
\\
$^{1}$Instituto de Astronom\'ia, Universidad Nacional Aut\'onoma de  M\'exico, A.~P. 70-264, C.P. 04510, M\'exico, D.F., Mexico \\
$^{2}$Department of Physics \& Astronomy, Johns Hopkins University, Bloomberg Center, 3400 N. Charles St., Baltimore, MD 21218, USA\\
$^{3}$Instituto de Astrof\'isica de Andaluc\'ia (IAA/CSIC), Glorieta de la Astronom\'{\i}a s/n Aptdo. 3004, E-18080 Granada, Spain \\
$^{4}$Departamento de F\'{\i}sica Te\'orica y del Cosmos, University of Granada, Facultad de Ciencias (Edificio Mecenas), E-18071 Granada \\
$^5$Leibniz-Institut f\"ur Astrophysik Potsdam (AIP), An der Sternwarte 16, D-14482 Potsdam, Germany\\
$^{6}$Department of Physics, Institute for Astronomy, ETH Z\"urich, CH-8093 Z\"urich, Switzerland\\
$^{7}$PITT PACC, Department of Physics and Astronomy, University of Pittsburgh, Pittsburgh, PA 15260, USA \\
$^{8}$Sydney Institute for Astronomy, School of Physics, University of Sydney, NSW 2006, Australia\\
$^{9}$INAF-Osservatorio Astrofisico di Arcetri - Largo Enrico Fermi, 5 - I-50125 Firenze, Italy\\
$^{10}$Australian Astronomical Observatory, PO Box 915, North Ryde, NSW 1670, Australia\\
$^{11}$Department of Physics and Astronomy, Macquarie University, NSW 2109, Australia\\
$^{12}$University of Vienna, Department of Astrophysics, T\"urkenschanzstr. 17, 1180 Vienna, Austria
}

\date{Accepted XXX. Received YYY; in original form ZZZ}

\pubyear{2016}

\hypersetup{draft}
\begin{document}
\label{firstpage}
\pagerange{\pageref{firstpage}--\pageref{lastpage}}
\maketitle

\begin{abstract}

We present an updated version of the mass--metallicity relation (MZR) using integral field spectroscopy data obtained from 734 galaxies observed by the CALIFA survey. These unparalleled spatially resolved spectroscopic data allow us to determine the metallicity at the same physical scale ($\mathrm{R_{e}}$) for different calibrators. We obtain MZ relations with similar shapes for all calibrators, once the scale factors among them are taken into account. We do not find any significant secondary relation of the MZR with either the star formation rate (SFR) or the  specific SFR for any of the calibrators used in this study, based on the analysis of the residuals of the best fitted relation. However we do see a hint for a (s)SFR-dependent deviation of the MZ-relation at low masses (M$<$10$^{9.5}$M$_\odot$), where our sample is not complete. We are thus unable to confirm the results by \citet{mann10}, although we cannot exclude that this result is due to the differences in the analysed datasets. In contrast, our results are inconsistent with the results by \citet{lara10a}, and we can exclude the presence of a SFR-Mass-Oxygen abundance Fundamental Plane. These results agree with previous findings suggesting that either (1) the secondary relation with the SFR could be induced by an aperture effect in single fiber/aperture spectroscopic surveys, (2) it could be related to a local effect confined to the central regions of galaxies, or (3) it is just restricted to the low-mass regime, or a combination of the three effects. 
\\
\\
\end{abstract}


\section{Introduction}

Galaxies in the Local Universe are perfect laboratories for the study
of the star formation and chemical enrichment history. Their
spectroscopic properties retain the fossil records of cosmological
evolution. For that reason, they possess correlations between their
different properties that are a consequence of that evolution, like
the so-called star-formation main sequence \citep[SFMS, e.g.][]{brin04} or the
mass-metallicity (MZ) relations. Those correlations change
with cosmological time tracing the evolution of the stellar
populations \citep[e.g.][]{dave11}.


The MZ relation determined from emission-line diagnostics was formally
presented by \citep[][T04 hereafter]{tremonti04}, although
the correlation appears in different forms in the earlier literature
going back several decades \citep[e.g.]{leque79,garn87,1992MNRAS.259..121V}. The MZ
relation exhibits a strong correlation between the stellar mass and
the average oxygen abundance in galaxies. Derived for a few tens of
thousand of galaxies observed by the SDSS survey, it extends over 4
order of magnitudes in mass and presents a small dispersion
($\sigma_{log(O/H)}\sim$0.1 dex).  This correlation has been confirmed
at different redshifts, showing a clear evolution with cosmological
time, as a result of the increase of the stellar-mass and the oxygen
enrichment
\citep[e.g.][]{erb06,erb08,henry13,saviane14,maier14,maier15,salim15,maier16}. Its
functional form may depend on the adopted abundance calibrator
\citep[e.g.,][]{2008ApJ...681.1183K}. However, it is rather stable
when using single aperture spectroscopic data or spatial resolved
information \citep[e.g.,][]{2012ApJ...756L..31R,sanchez14}. Despite the differences in methodology and
physical interpretation, a similar relation has been reported for the
metallicity of stellar populations
\citep[e.g.][]{gallazzi+05,rosa14b}.

The original interpretation of the shape of this relation (T04), with
its clear saturation in the abundance at high stellar mass, was that
galactic outflows regulate the metal content. The idea here is that
outflows are more efficient for larger star formation rates presumed
to occur in higher mass galaxies. This interpretation was re-phrased
recently by \cite{belf16a} indicating that galaxies reach an
equilibrium metallicity in which the metals expelled by outflows are
compensated by those produced by star-formation. However, this
interpretation requires that the outflows are strong enough to escape
the gravitational potential and expel a substantial fraction of the
generated oxygen, overcoming the effect of ``rainfall'' and metal
mixing. \cite{2012ApJ...756L..31R} presented an alternative
explanation in which the effect of outflows is not required. They show
that the integrated relation is easily derived from a new, {\it more
  fundamental} relation between the stellar mass density and the local
oxygen abundance. This relation was confirmed with larger statistics
by \cite{sanchez13} and more recently by \cite{2016MNRAS.463.2513B},
using MaNGA data \citep{manga}. Under the proposed scenario the
stellar mass growth and the metal enrichment are both dominated by
local processes, basically the in-situ star formation, with little
influence of outflows or radial migrations. The differential
star-formation history from the inner to the outer regions, known as
local downsizing \citep[e.g.][]{perez13,ibarra16}, and the fact that
oxygen enrichment is directly coupled with star-formation explain both
the local and the global MZ relations naturally. Moreover, the local
relation explains the oxygen abundance gradients observed in galaxies
\citep[e.g.,][]{sanchez14b}, as recently shown by
\cite{2016MNRAS.463.2513B}. Under this assumption the plateu reached
in the MZ relation is a pure consequence of the maximum yield of
oxygen abundance and a characteristic depletion time, as already
suggested by \citet{pily07}.

On the other hand, \cite{mann10} and \cite{lara10a} presented almost simultaneously an analysis of the dependence of the MZ relation with the SFR, that they called Fundamental Mass-Metallicity relation (FMR), in the first case, and star-formation-Mass-Oxygen abundance Fundamental Plane (FP), in the second case. They both showed that there is a secondary relation in the sense that, at a fixed stellar mass, galaxies with stronger SFR exhibit lower oxygen abundances. Although the adopted functional form for this secondary relation was different in both studies the conclusions were very similar. That correlation is a bit anti-intuitive, since oxygen abundance is enhanced due to star-formation. Both studies were based on two similar sub-samples of the same observational dataset. They both used the SDSS spectroscopic survey at z$\sim$0.1 (from which the oxygen abundance and the star-formation were derived) combined with the photometric information to derive the integrated stellar mass. They applied aperture corrections to the SFR \citep{brin04}, due to the strong aperture effects of the SDSS single fiber spectroscopic information \citep[e.g.][]{2013A&A...553L...7I,2016A&A...586A..22G}. However,  they did not apply any aperture correction to oxygen abundance indicators, what could be substantially important \citep[e.g.][]{2016ApJ...826...71I}. Even more, the applied corrections depend on certain correlations between the SFR and the color gradients in galaxies, that are not fully tested. Recent results indicate that those corrections could be strongly affected by the assumed correlations \citep{duarte16}.

This result is under discussion. \cite{sanchez13} showed that using
catalog of HII regions extracted from the CALIFA dataset
\citep{sanchez12a} observed up to that date (150 galaxies), the
secondary relation cannot be confirmed. In a contemporary article,
\cite{2013A&A...550A.115H} shown that using drift-scan integrated
spectra the secondary relation is not present. Indeed,
\cite{2012ApJ...756L..31R} had already shown that the relation with
the specific star-formation rate (sSFR, in the form of the EW(\Ha)) of
the local MZ relation does not present a secondary trend, following
the primary relation between the SFR and the mass, as studied in
detail in \cite{sanchez13}. Actually, T04 explored the residuals of
the MZ relation, and found that there was no evidence for a relation
with the EW(\Ha). A different approach was presented by
\cite{2014ApJ...797..126S}. In this case they analyzed the SDSS data
exploring the relation between the oxygen abundance and the sSFR for
different mass bins. They found a clear anticorrelation but
considerably weaker than the one presented by \citet{mann10} and
\cite{lara10a}. They repeated the analysis using the CALIFA data
presented by \cite{sanchez13}, finding a similar result (i.e., that
there is a secondary correlation). However, those correlations could
be easily explained as a consequence of the primary SFR-Mass and
Mass-Z relations, and they disappear if the dependence with the Mass
is removed.

{ Furthermore}, \cite{2012ApJ...745...66M} showed that the secondary relation
is not seen in their data, and they proposed a secondary relation with
the gas fraction as previously explored by
\cite{1992MNRAS.259..121V}. In the same line, \citet{bothwell16} shows
that the primary driver for the FMR is the relation between SFR and
molecular gas \citep{kennicutt89}, in their analysis of a sample
covering a wide range of redshifts ($0<z<2$). { They found a clear
  trend between the residual of the MZ-relation and the molecular gas
  mass (Fig. 4 of that article).} { However, in a previous article
  exploring galaxies in a much narrower redshift range at the Local
  Universe \cite{both13} shown that while there is secondary relation
  of the MZR with the atomic cold-gas (HI), they cannot confirm the
  existence of a secondary relation with the molecular gas (H2). Since
  atomic gas is not a tracer of the SFR that secondary relation would
  imply a new, different relation than the proposed FMR, in agreement
  with the results presented by \cite{2012ApJ...745...66M}, described
  before. Therefore, the presence of secondary relation with the
  molecular gas mass, the tracer of the SFR, is under discussion, even
  more if it is considered the strong correlation between the CO/H2
  correction factor with the metallicity \citep[e.g.][]{bola13}, as
  already pointed out by \cite{both13}. }

Furthermore, recent results have shown that even using single aperture spectroscopic data the secondary relation between the MZ and the SFR may disappear when using particular abundance calibrators \citep[e.g.][]{kash16}. And, in any case, it seems to be weaker than what it was previously reported \citep[e.g.][]{telf16}, and strongly dependent on the assumptions behind the derivation of the three parameters involved: for example, the SFR derivation is based on calibrations that assume solar abundances \citep[e.g.][]{kennicutt89}. Finally, in a complementary study presented by Barrera-Ballesteros et al. (submitted), it is not found any  secondary correlation with the SFR based on the analysis of 
the sample of galaxies observed by the MaNGA survey \citep{manga} up to date.

In the current article we revisit the MZ relation and its possible dependence with the SFR using the integral field spectroscopic data provided by the full sample of galaxies observed by the CALIFA survey \citep{sanchez12a}. The distribution of the article is as follows: in Section \ref{sample} we present the sample of galaxies and the adopted dataset; the adopted calibrators to derive the oxygen abundance are described in Section \ref{sec:calibrators}; in Section \ref{sec:MZ} we present the Mass-Metallicity (MZ) relation derived using these data, and the possible dependence with the SFR is explored in Section \ref{sec:residuals}; a comparison with the so-called Fundamental Mass-Metallicity relation is included in Section \ref{sec:FMR}, and the possible dependence of the residuals of the MZ relation with either the sSFR or the residual of the SF Main Sequence are explored in Section \ref{sec:salim}; finally the results are discussed in Section \ref{sec:con}.

\section{Sample and data}
\label{sample}

The analyzed sample comprises all the galaxies with good quality spectroscopic data observed with the low resolution setup (V500) by the CALIFA survey   \citep[][]{sanchez12a} and by a number of CALIFA-extensions listed in \citet{DR3} up to October 9th 2016. It includes the 667 galaxies from the 3rd CALIFA Data Release \citep{DR3}, and in addition we include those galaxies with good quality data excluded from DR3 because either they did not have SDSS-DR7 imaging data (a primary selection for DR3) or they were observed after the final sample was closed (i.e., after November 2015). The final sample comprises a total of 734 galaxies. 

\begin{figure}
 \includegraphics[width=8.8cm]{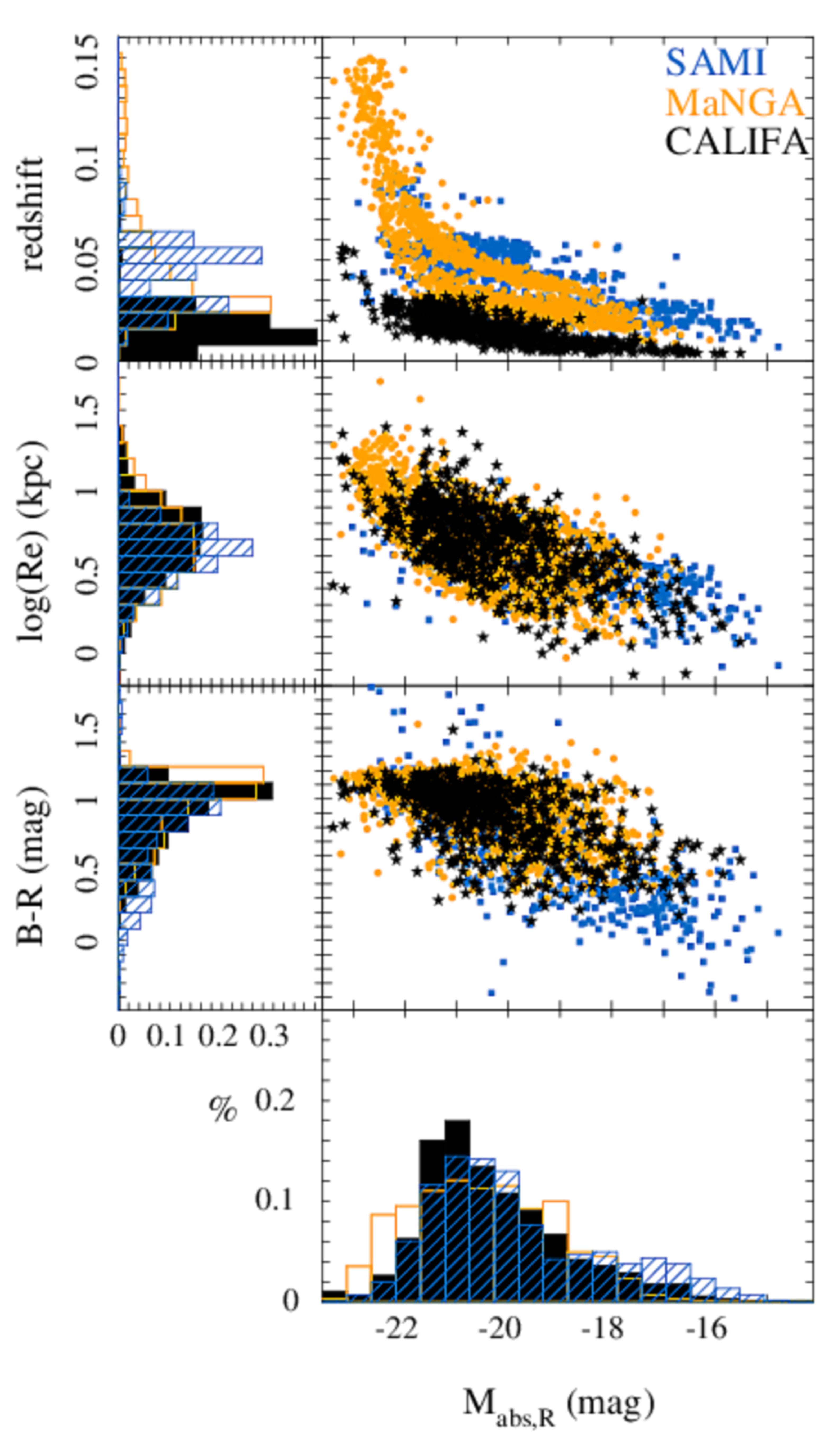}
\caption{Comparison of the main properties of the analyzed sample (CALIFA) with that of similar samples observed by current on-going IFU surveys (MaNGA, SAMI). The plot shows the redshift, effective radius and color distributions along the absolute magnitudes, together with the corresponding frequency histograms for the current analysed sample (black stars), the MaNGA sample distributed in the SDSS-DR13 (MPL-4 sample), and the SAMI sample observed up to summer 2016.} 
 \label{fig:sample}
\end{figure}

The main properties of this sample are shown in Figure \ref{fig:sample}, including the redshift, effective radius, and $B-R$ color distribution along the absolute magnitude of the galaxies. All the parameters were derived directly from the datacubes. In the case of the redshift was derived as part of the Pipe3D analysis described later. The photometric parameters were extracted from the datacubes by convolving the Johnson filter responses and applying the zero-points listed in \citet{fukugita95}. For comparison purposes we have included the same properties for the galaxies observed by the MaNGA \citep{manga} and SAMI \citep{sami} surveys. For clarifying purposes and prior to any further analysis, we find interesting to include this comparison due to the similar goals of the three surveys, that can lead to the analysis of the current topic using the three datasets (already performed by Barrera-Ballesteros et al. submitted with MaNGA data. For MaNGA we used the published sample included in the SDSS-DR13 \citep[the so-called MPL-4 dataset,][]{SDSS_DR13}. For SAMI we used the catalog of galaxies observed up to summer 2016. The figure shows that the analyzed sample covers a much narrower range of redshifts, offers a better physical resolution per galaxy and covers a similar range of absolute magnitudes, colors and effective radius, than that of the other two major IFU surveys. On one hand, MaNGA presents a flatter distribution in masses by construction, at the cost of observing the most massive ones at a considerably larger redshift (and lower physical resolution). On the other hand, SAMI samples better the low luminosity/bluer range of the color-magnitude diagram. This is already known for the CALIFA Mother Sample, that it is not complete below 10$^{9-9.5}$M$_\odot$ \citep[e.g.][]{walcher14}. However, the analyzed sample is more similar to the CALIFA DR3 sample \citep{DR3}, covering better the lower mass/luminosity range of the color-magnitude diagram, and being more suitable for the proposed exploration, due to the inclusion of the CALIFA-extended samples.

The details of the CALIFA survey, including the observational strategy and data reduction are explained in \citet{sanchez12a} and \citet{DR3}. All galaxies were
observed using PMAS \citep{roth05} in the PPaK configuration
\citep{kelz06}, covering an hexagonal field of view (FoV) of
74$\arcsec$$\times$64$\arcsec$, which is sufficient to map the full optical extent of the galaxies up to two to three disk effective radii. This is possible because of the diameter selection of the CALIFA sample \citep{walcher14}. The observing strategy guarantees complete
coverage of the FoV, with a final spatial resolution of
FWHM$\sim$2.5$\arcsec$, corresponding to $\sim$1 kpc at the average redshift of the survey \citep[e.g][]{rgb15,DR3}. The sampled wavelength range and
spectroscopic resolution for the adopted setup (3745-7500 \AA,
$\lambda/\Delta\lambda\sim$850, V500 setup) are more than sufficient
to explore the most prominent ionized gas emission lines from
[Oii]$\lambda$3727 to [Sii]$\lambda$6731 at the
redshift of our targets, on one hand, and to deblend and subtract the underlying stellar population, on the other
\citep[e.g.,][]{kehrig12,cid-fernandes13,cid-fernandes14,sanchez13,sanchez14,Pipe3D_I}
In addition, most of the objects are observed using a higher resolution setup, covering only the blue end of the spectral range (3700-4800\AA,
$\lambda/\Delta\lambda\sim$1650, V1200 setup), that it is not used in the current analysis. The current dataset was
reduced using version 2.2 of the CALIFA pipeline, whose modifications with respect to the previous ones \citep{sanchez12a,husemann13,rgb15} are described in \citet{DR3}. The final dataproduct of the reduction is a datacube comprising the spatial information in the {\it x} and {\it y} axis, and the spectral one in the {\it z} one. For further details of the adopted dataformat and the quality of the data consult \citet{DR3}.

\section{Analysis}
\label{sec:ana}

We analyze the datacubes using the {\sc Pipe3D} pipeline \citep{Pipe3D_II}, which is designed to fit the continuum with stellar population models and measure the nebular emission lines of IFS data. This pipeline is based on the {\sc FIT3D } fitting package \citep{Pipe3D_I}. The current implementation of {\sc Pipe3D} adopts the GSD156 library of simple stellar populations \citep{cid-fernandes13}, that
comprises 156 templates covering 39 stellar ages (from 1Myr to 13Gyr), and 4 metallicities (Z/Z$_\odot$=0.2, 0.4, 1, and 1.5). This templates have been extensively used within the CALIFA collaboration \citep[e.g.][]{perez13,rosa14}, and for other surveys \citep[e.g.][]{ibarra16}. Details of the fitting procedure, dust attenuation curve, and uncertainties of the processing of the stellar populations are given in \citet{Pipe3D_I,Pipe3D_II}. 

In summary, for the stellar population analysis it is performed a spatial binning to each datacube to reach a
goal S/N of 50 across the FoV. Then, the
stellar population fitting was applied to the coadded spectra within each spatial bin. Finally, following the procedures described in \citet{cid-fernandes13} and \citet{Pipe3D_I}, we estimate the
stellar-population model for each spaxel by re-scaling the best fitted model within each spatial bin to the continuum flux intensity in the corresponding spaxel. This model is used to derive the stellar mass density at each position, in a similar way as described in \citet{mariana16}, adopting the Salpeter IMF \citep{Salpeter:1955p3438}, and then coadded to estimate the integrated stellar mass of the galaxies. That estimation of the stellar mass has a typical error of 0.15 dex, as described in \citet{Pipe3D_II}.

The stellar-population model spectra are then substracted to the original cube to create a gas-pure cube comprising only the ionised gas emission lines (and the noise). Individual emission line fluxes were then measured spaxel by spaxel using both a single Gaussian fitting for each emission line and spectrum, and a weighted momentum analysis, as described in \citet{Pipe3D_II}. For this particular dataset we extracted the flux intensity of the following emission lines: \Ha, \Hb, \oii\ \lam3727,
\oiii\ \lam4959, \oiii\ \lam5007, \nii\ \lam6548,
\nii\ \lam6583, \sii \lam6717 and \sii \lam6731. The intensity maps for each of these lines are corrected by dust attenuation, derived using
the spaxel-to-spaxel \Ha/\Hb\, ratio. Then it is assumed a canonical value of 2.86 for this ratio \citep{osterbrock89}, and adopting a \citet{cardelli89} extinction law and R$_{\rm V}$=3.1 \citep[i.e., a Milky-Way like extinction law][]{schlegel98}.

The spatial resolved oxygen abundance we select only those spaxels which ionization is clearly compatible with being produced by star-forming areas following \citet{sanchez13}. For doing so we select those spaxel located below the \citet{kewley01} demarcation curve in the classical BPT diagnostic diagram \citep[\oiii/\Hb\, vs \nii/\Ha\, diagram]{baldwin81}, and 
with a EW(\Ha) larger than 6\,\AA. These criteria ensure that the ionization is compatible with being due to young stars \citep{sanchez14}, and therefore the abundance calibrators can be applied. The \Ha\, luminosity is derived by correcting for the cosmological distance the dust corrected \Ha intensity maps. Then, by applying the \citet{kennicutt98} calibration (for the Salpeter IMF), we derive the spatial resolved distribution of the SFR surface density, and finally the integrated SFR. We did not apply the very restrictive selection criterium indicated before (EW(\Ha)$>$6\,\AA) for the derivation of the SFR in order to include the diffuse ionized gas. While in retired galaxies (or areas within galaxies) this ionized gas is most probably dominated by post-AGB ionization \citep[e.g.][]{sarzi10,sign13,2016A&A...586A..22G}, in star-forming galaxies the photon leaking from \ION{H}{ii} regions may represent a large contribution to the integrated \Ha\, luminosity \citep[e.g.][]{rela12,mori16}, and the SFR estimation. The contamination of the post-AGB ionization in our derivation of SFR represents a contribution more than 2 orders of magnitude lower than the actual SFR for galaxies located in the star-formation main sequence \citep[e.g.][]{catalan15,mariana16,duarte16}. Thus, it affects the SFR by less than a $\sim$1\% for those galaxies. For \Ha\, we applied a signal-to-noise cut of 3$\sigma$ spaxel-by-spaxel, while for the remaining lines we relax that cut down to 1$\sigma$. The cut in \Ha\, ensures a positive detection of the ionized gas, while the cut in the other lines limits the error for the derived parameters.

\begin{figure}
 \minipage{0.49\textwidth}
 \includegraphics[clip,trim=100 30 120 100, width=\linewidth]{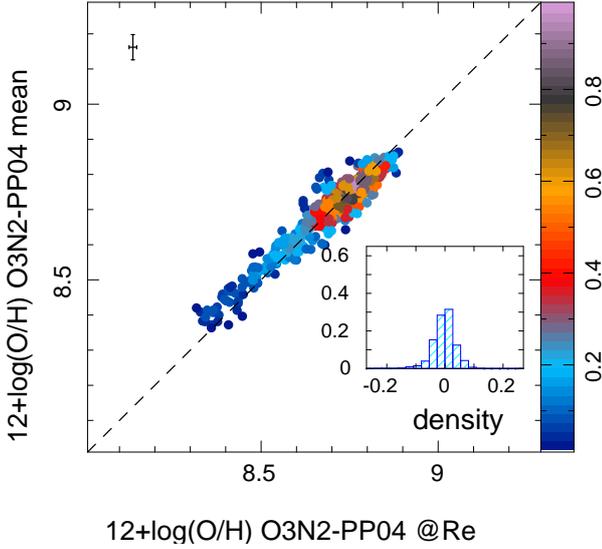} 
 \endminipage
\caption{ Comparison between the average oxygen abundance across the entire Field-of-View of the datacubes (beyond $\sim$2.5 r$_e$) and the characteristic oxygen abundance at the effective radius using the linear regression technique described in the text. Each solid circle corresponds to an individual galaxy, with a color code indicating the density of points normalized to the peak density. The upper left error-bar indicates the median error estimated for each parameter. The inset shows a histogram of the difference between the two estimations of the oxygen abundance. }
 \label{fig:com_OH_PP04}
\end{figure}

Finally, we determine the oxygen abundance for each spaxel using the different calibrators described in Sec.\ref{sec:calibrators}, using the dust extinction corrected intensity maps for the set of emission lines described before, based on the dust extinction correction described before. We adopted as a characteristic oxygen abundance for each galaxy the value derived at the effective radius $\rm{R_{e}}$. This abundance match pretty well with the average abundance across the optical extension of the galaxies, as demonstrated by \citet{sanchez13}. To derive this abundance we perform a linear fitting to the deprojected abundance gradient within a range of galactocentric distances between 0.5 and 2.0 $\rm{R_{e}}$, as described in \cite{sanchez13, 2016A&A...587A..70S}. This abundance gradient has a nominal error well below the typical error for a single aperture spectroscopic derivation \citep[][]{2014ApJ...797..126S} due to the larger number of sampled points for each single galaxy, being typically $\sim$0.03 dex \citep[e.g.][]{sanchez13}. The photometric properties of the galaxies (stellar mass [M$_{\odot}$], PA and ellipticity) were obtained from CALIFA DR3 tables\footnote{\url{http://califa.caha.es/DR3}}, derived as described in \citet{walcher14}. We use as final sample those galaxies where it is possible to determine the oxygen abundance at the $\rm{R_{e}}$ fulfilling the above criteria (612 objects).  { Figure \ref{fig:com_OH_PP04} shows the comparison between the average oxygen abundance derived using all the suitable spaxels within the FoV of the datacubes (i.e., those ones compatible with being ionized by star-formation), and the characteristic abundance derived at the effective radius for a particular oxygen abundance calibrator of the ones described in Sec. \ref{sec:calibrators} (a similar comparison for the remaining calibrators is included in Appendix \ref{sec:OH_comp}. As already noticed by \citet{sanchez13}, the characteristic abundance is very good representation of the average abundance of a galaxy, with a correlation following the one-to-one relation, and dispersion of $\sim$0.03 dex. However, the typical error for the characteristic abundance is, on average, a factor two lower than that of the mean oxygen abundance, as shown in Fig. \ref{fig:com_OH_PP04}. }

\begin{figure*}
 \minipage{0.99\textwidth}
 \includegraphics[width=\linewidth]{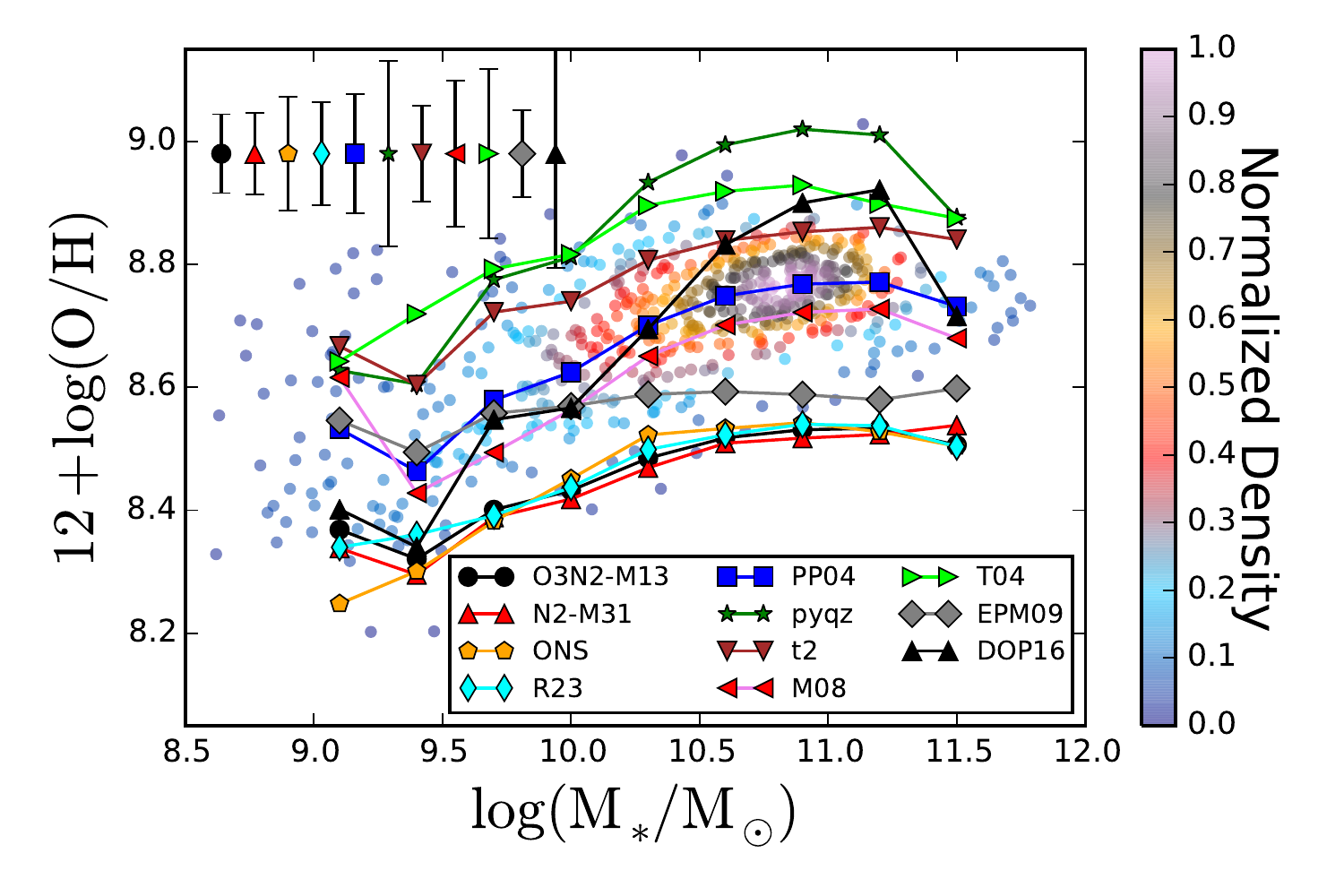} 
 \endminipage
\caption{ Mass-metallicity relation derived using different oxygen abundance calibrators for the 612 galaxies extracted from the final CALIFA sample analyzed in this article, for the abundances measured at the effective radius (Reff). Colored solid symbols indicate the distribution for the individual galaxies when adopting the PP04 calibrator, with the color-code indicating the density of points (magenta indicating larger density, and light-blue indicating lower density). Line-connected symbols represent median values at a given mass bin for the different calibrators, as indicated in the inset. The error bars in the top-left symbols represent the average standard deviation for each indicator with respect to the median value at different mass bins. A description of the different calibrators is included in Sec. \ref{sec:calibrators}.} 
 \label{fig:MZ}
\end{figure*}

\subsection{Abundance Calibrators}
\label{sec:calibrators}

Accurate abundance measurements for the ionized gas in galaxies can be
derived from metal recombination lines without requiring to know the
electron temperature (T$_e$), since they have a similar temperature
dependence to the hydrogen recombination lines
\citep[e.g.][]{2007ApJ...656..186B}. A major advantage of this method
is that they are also insensitive to the temperature fluctuations,
characterized by the so-called $t_2$ parameter
\citep[e.g.][]{odell03}. However those lines are extremely weak or
they are not accessible in the optical wavelength range, in most of
the cases \citep{2007ApJ...656..186B}, in particular for
low-resolution spectroscopy ($R\sim1000$).

A slightly better approach is to determine the electron temperature
(T$_e$) from the ratio of auroral to nebular line intensities, such as
[\ION{O}{iii}]~$\lambda$4363/[\ION{O}{iii}]~$\lambda\lambda$4959,5007
\citep{osterbrock89}. However, this procedure, known as the "Direct
Method", can be affected by electron temperature fluctuations. Even
more, as the metallicity increases, the electron temperature decreases
due to the cooling by metals and the auroral lines eventually become
too faint to measure \citep[e.g.,][]{marino13}. This same effect makes
the strongest oxygen line ratio, $R23$ =
([\ION{O}{ii}]~$\lambda$3727]+[\ION{O}{iii}]~$\lambda$4959,5007)/\Hb),
  to be bi-valuated.

It is still under debate which of both procedures derive a more
accurate estimation of the oxygen abundance. While it is broadly
accepted that the temperature fluctuations would affect our
measurements based on the second method \citep[e.g.][]{2006RMxAC..26R.163P},
recent results indicate that the oxygen abundances derived using it
agree much better with the value derived from high precision stellar
spectroscopy \citep{2016ApJ...830...64B}. In any case both methods
requires high accurate measurements of the emission line intensity of
extremely faint emission lines.

\begin{table*}
\caption[Fitting parameters for the MZR and its scatter.]{Fitting parameters for the MZR and its scatter for the set of abundance calibrators used in this study. For each calibrator we list: the standard deviation of the original distribution of the oxygen abundances ($\sigma_{log(O/H)}$); the parameters $a$ and $b$ from the fitting of Eq.\ref{eq:fit} to the MZR;  $\sigma$ MZ-res lists the standard deviation of the residuals after subtracting the best fit to the MZR; the parameters $\alpha$ and $\beta$ represent the linear fitting of the residuals of the MZR respect to the SFR (see Sec.\ref{sec:residuals}); $\sigma \Delta$ MZ-res lists the standard deviation of the residuals of the linear fitting using the above parameters. We included the third decimal in the $\sigma$ to highlight any possible difference, however below the 2nd decimal it is totally insignificant.}
\label{tab:val}
\begin{tabular} {c c c c c r r c }
\hline
Metallicity   & $\sigma_{log(O/H)}$& \multicolumn{2}{c}{MZ Best Fit} & $\sigma$ MZ-res & \multicolumn{2}{c}{$\Delta$MZ Best Fit } &  $\sigma$ $\Delta$MZ-res  \\
 Indicator & (dex) & $a$ (dex)  & $b$ (dex $/ \log(M_{\odot}))$ & (dex) & $\alpha$ (dex)  & $\beta$ (dex/$\log(M_{\odot} \, \mathrm{yr^{-1}}))$ & (dex) \\
\cline{3-4}
\cline{6-7}
O3N2-M13 & 0.077 & 8.53 $\pm$ 0.04 & 0.003 $\pm$ 0.037 & 0.060 & -0.02 $\pm$ 0.01 & -0.007 $\pm$ 0.005 & 0.061 \\
PP04     & 0.111 & 8.76 $\pm$ 0.06 & 0.005 $\pm$ 0.037 & 0.087 & -0.02 $\pm$ 0.01 & -0.011 $\pm$ 0.007 & 0.088 \\
N2-M13   & 0.080 & 8.53 $\pm$ 0.04 & 0.004 $\pm$ 0.023 & 0.060 & -0.01 $\pm$ 0.01 & -0.013 $\pm$ 0.005 & 0.058 \\
ONS      & 0.095 & 8.55 $\pm$ 0.04 & 0.006 $\pm$ 0.023 & 0.082 & -0.01 $\pm$ 0.01 & -0.004 $\pm$ 0.008 & 0.082 \\
R23      & 0.075 & 8.54 $\pm$ 0.03 & 0.003 $\pm$ 0.020 & 0.065 & -0.01 $\pm$ 0.01 & -0.014 $\pm$ 0.009 & 0.064 \\
pyqz     & 0.183 & 9.00 $\pm$ 0.12 & 0.007 $\pm$ 0.061 & 0.147 &  0.01 $\pm$ 0.01 &  0.086 $\pm$ 0.015 & 0.145 \\
t2       & 0.076 & 8.85 $\pm$ 0.01 & 0.007 $\pm$ 0.001 & 0.064 &  0.00 $\pm$ 0.00 &  0.006 $\pm$ 0.004 & 0.063 \\
M08      & 0.107 & 8.72 $\pm$ 0.10 & 0.004 $\pm$ 0.057 & 0.087 & -0.01 $\pm$ 0.01 &  0.003 $\pm$ 0.001 & 0.087 \\
T04      & 0.145 & 8.92 $\pm$ 0.04 & 0.008 $\pm$ 0.029 & 0.133 &  0.01 $\pm$ 0.01 &  0.014 $\pm$ 0.016 & 0.133 \\
EPM09    & 0.062 & 8.59 $\pm$ 0.03 & 0.001 $\pm$ 0.017 & 0.060 & -0.01 $\pm$ 0.01 & -0.001 $\pm$ 0.006 & 0.060 \\
DOP16    & 0.249 & 8.86 $\pm$ 0.19 & 0.008 $\pm$ 0.094 & 0.183 & -0.01 $\pm$ 0.02 &  0.041 $\pm$ 0.023 & 0.186 \\
\hline
\end{tabular}
\end{table*}

For this reason, in most of the cases they are used calibrators
based on strong emission lines, first proposed by \citet{pagel79} and \citet{allo79} \citep[see][for a review]{angel12}. All those calibrators attempt to derive the oxygen (and nitrogen in some cases) abundance based on a known relation between a particular strong-line ratio (or a set of them) and the required abundance. Such relations could be derived using two different approaches: (i) by comparing the known abundances of a set of \ION{H}{ii} regions with the measured line ratio (or ratios), or (ii) by comparing the line ratios predicted by photo-ionization models for a set of modeled \ION{H}{ii} regions, with the input abundances included in those models. Both procedures produce two different families of abundance calibrators, those anchored to the ``Direct Method" and those anchored to photoionization models. It is a well known and long standing issue that both families of abundances calibrators derive different results \citep[e.g.][]{kehrig08,mori16}. In general, the former ones derive abundances ~0.2-0.3 dex lower than the later ones, for the larger abundance values \citep[see][for a counter example]{epm14}.

There is a long standing discussion on the nature of this
discrepancy. Those supporting the "Direct Method" approach claim that
it is the method with less number of assumptions, being based on basic
atomic physics and very few assumptions on the ionization conditions
\citep[e.g.][]{pilyugin10}. They criticize the photo-ionization models
for two main reasons: (i) they are based on strong assumptions on the
physical behavior of atmospheres of the ionizing population (largely
unknown), and the structure of ionized nebulae (shape, density
distribution...), and (ii) they cannot derive the oxygen abundance
without assuming strong correlations, or trends, between this
parameter and the ionization strength \citep[e.g.][]{epm14}, and/or
the N/O and S/O relative abundances
\citep[e.g.][]{kewley02,dopita13}. In some cases those relations are
not imposed, but they are implicit, despite the fact that they are no
introduced as a prior in the derivation of the abundances
\citep[e.g.][]{blanc15}. The reason is the strong relation between
metal abundance, star effective temperature, and blanketing, and
between them and the ionization strength that implies a correlation
even if it is not imposed. That correlation was already known or
hinted since decades \citep[e.g.][]{evans85}, and it was recently
revisited by \citet{sanchez14b}. In summary, by adopting a particular
library of ionizing stars it is assumed a particular correlation
between oxygen abundance and ionization strength
\citep[e.g.][]{pepe88,mori16}.

On the other hand, those supporting the photo-ionization model calibrators claim that the abundances derived by the direct method are far too low for being real \citep[e.g.][]{blanc15}, since they can hardly reproduce super-solar oxygen abundances. It is frequently claimed that the fact that the abundances estimated using 
photoionization models are in a better agreement with the values measured using recombination lines is an indication that it is more accurate to use photoionization models in this regime \citep[e.g.][]{maio08}. That claim has lead to mixed calibrators, like the one presented by \citet{pettini04} or \citet{maio08}. Those ones anchor the abundances below 12+log(O/H)$\sim$8.3 to estimations based on the Direct Method and above that value to estimations based on the photoionization models. However, the argument has a logic flow, since the discrepancy between the measurements based on recombination lines and those based on the direct method are supposed to be due to inhomogeneties in the electron temperature, which in general are not included in photoionization models (a priori). Thus, if the values derived using the Direct Method should be corrected by the $t_2$ effect then, the values derived by photo-ionization models should be corrected too, since they predict very low values for the $t_2$ \citep{2006RMxAC..26R.163P}. Therefore the discrepancy with the values estimated using recombination lines holds, but in the opposite direction.

In order to minimize the effects of selecting a particular abundance calibrator on biasing our results and explore in the most general way the shape of the MZ relation we have not restricted our analysis to a single calibrator. We derive the abundance using (i) calibrators anchored to the ``direct method", including the O3N2 and N2 calibration proposed by \citet[][O3N2-M13 and N2-M13 hereafter]{marino13}, the R23 calibration proposed by \citet{2004ApJ...617..240K} as described in \citet{rosales11}, modified to anchor the abundances to the direct method (R23 hereafter, S\'anchez et al., in prep.), the calibrator proposed by \citet{pilyugin10} (ONS hereafter), and a modified version of O3N2 that includes the effects of the nitrogen-to-oxygen relative abundance proposed by \citet{epm09} (EPM09 hereafter); (ii) a $t_2$ correction proposed by \cite{2012ApJ...756L..14P} for an average of the abundances derived using the four previous methods, that produce in general very similar results within the nominal errors ({\sc t2} hereafter); (iii) two mixed calibrators, based on the O3N2 indicator \citep[][PP04 hereafter]{pettini04}, and the R23 indicator \citep[][M08 hereafter]{maio08}; and finally (iv) three calibrators based on pure photo-ionization models, the one included in the {\tt pyqz} code, that uses the O2, N2, S2, O3O2, O3N2, N2S2 and O3S2 line ratios as described in \citet{dopita13} (pyqz hereafter); a recent calibrator  proposed by \citet{dopita16cal} that uses just the N2/S2 and N2 line ratios (DOP16 hereafter); and finally the one adopted by \citet{tremonti04} in their exploration of the MZ relation based on the R23 line ratio (T04 hereafter). The complete list of calibrators is included in Table \ref{tab:val}.

This selection of calibrators is by no means complete, and it is clearly out of the scope of this article to analyzed the similarities and differences between them, both in the final estimated value for the abundances and in the physical assumptions behind them. With the current selection we try to cover a wide range of possible calibrators, mostly motivated to explore if there is a significant change on the main conclusions depending on the adopted one. { The complete list of characteristic oxygen abundances, stellar masses and star-formation rates are included in Appendix \ref{sec:dataset}}.

\section{Results}
\label{sec:results}

Once derived the integrated stellar masses, star-formation rates and the characteristic oxygen abundance for all the 612 galaxies, we explore the shape of the MZ relation and its possible dependence with the star-formation rate. We start by characterizing the shape of the MZ relation.

\subsection{The MZ relation}
\label{sec:MZ}

In Fig.\ref{fig:MZ} we plot the average abundances at different stellar mass bins for our set of calibrators, together with the individual values for the PP04 calibrator.  It is remarkable that almost all abundances follow a similar trend despite of the fact that the different calibrators are based on different assumptions and using different line ratios as indicators. Oxygen abundance increases with the stellar mass for $\rm M>10^{9.25}\,M_{\odot}$, reaching an asymptotic value for more massive galaxies, as described by \citet{pily07} \citep[the {\it equilibrium} value in the nomenclature of][]{belf15}. For masses below $\rm M<10^{9.25}\,M_{\odot}$ we have a large dispersion and little statistics. The absolute scale of the relation presents a dependence on the adopted calibrator. In general, calibrators based on photoionization models have a larger dynamical range what is reflected in the larger values for the standard deviation of the oxygen abundances shown in Tab. \ref{tab:val}, and higher abundances. In contrast, calibrators anchored to the direct method have a smaller dynamical range (thus, smaller standard deviations), with mixed calibrators lying in between. We also notice that the dispersion around the mean values for the different mass bins are considerably larger for calibrators based on photoionization models than for the rest. As expected, the {\sc t2} correction shifts the abundances based on the direct method toward values more similar to those derived using photoionization models, but at slightly lower abundances, in general, and a much smaller dispersion around the mean value for each mass bin.

\begin{figure*}
\minipage{0.495\textwidth}
\includegraphics[width=\columnwidth]{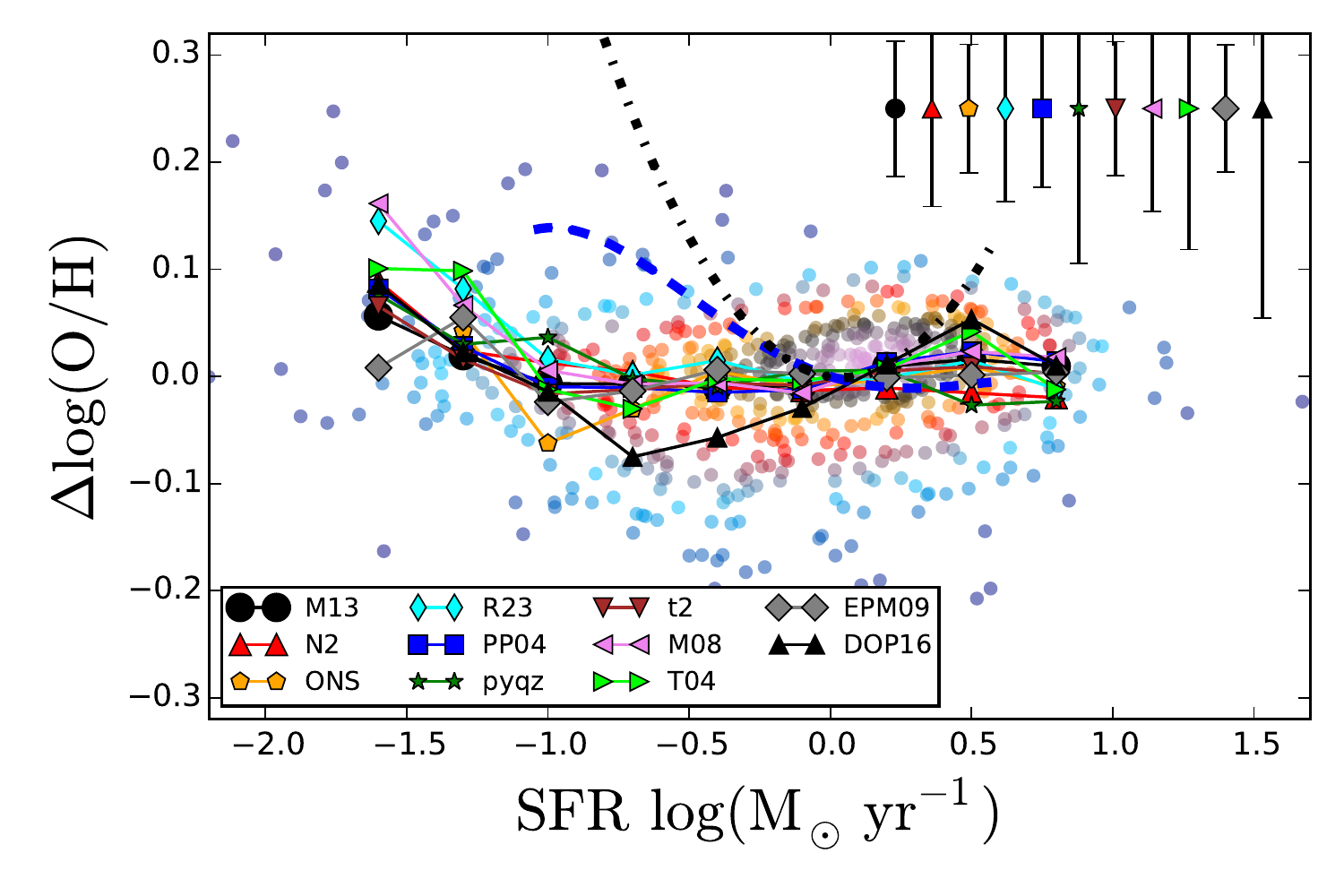}
\endminipage
\minipage{0.495\textwidth}
\includegraphics[width=\columnwidth]{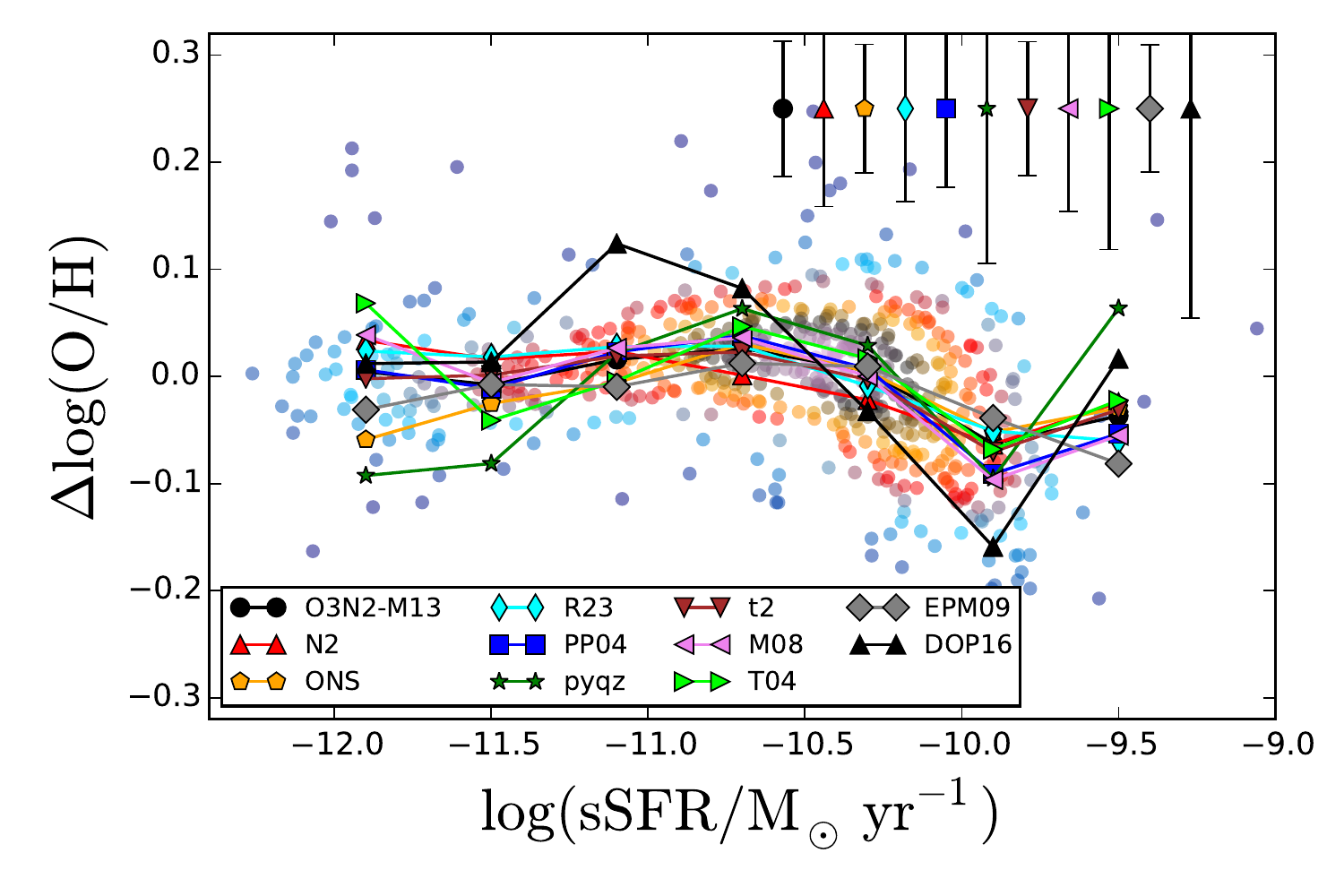}
\endminipage
\caption{Residuals of the MZ relations from the different analyzed calibrators against the SFR (left-panel) and the sSFR (right-panel). The colored solid circles corresponds to the values derived using the PP04 calibrator, and the line-connected symbols corresponds to median values for each bin and calibrator, following the notation presented in Fig. \ref{fig:MZ}. The error bars in the top-right represent the mean standard deviation of the residuals along the considered bins. The blue-dashed line represents the relation between the residuals and the SFR expected when using the secondary relation proposed by \citet{mann10}, and the black-dotted line represents the same relation when adopted the secondary relation proposed by \citet{lara10a}.} 
\label{fig:dMZ}
\end{figure*}

To characterize the MZ relation we fit the median values at different mass bins using the functional form between these two parameters introduced by \cite{mous11} and used by \cite{sanchez13}, for each calibrator:
\begin{equation}
y=a+b(x-c)\exp(-(x-c))
\label{eq:fit}
\end{equation}
where $y = 12+\log(\mathrm{O/H})$ and $x=\log(M_*/ M_{\odot})-8.0$.  This functional form has been motivated by the shape of the relation \citep{sanchez13}. The fitting coefficients, $a$, $b$ and $c$ represent the asymptotic metallicity, the curvature of the relation and the stellar mass where the metallicity reaches its asymptotic value. We fix $c = 3.5$ since for all the calibrators the asymptotic value is reached at $\log(M_*/M_{\odot})$ $\sim 11.5$.
This value was the one reported by \cite{sanchez13}, and therefore, by adopting it we can perform a direct comparison on the results. However, we should stress that using another value (like $c=4$, adopted by Barrera-Ballesteros, submitted), or fitting this parameter without any restriction does not change the main conclusions of this analysis. It will modify the numerical value of the $b$ parameter, but neither the general shape of the relation nor the dispersion around  this relation. In Tab.\ref{tab:val} we list the best-fitted parameters for the different calibrators. As expected from Fig.\ref{fig:MZ}, \temp-based calibrators show lower values of the asymptotic metallicity, in general. The curvature does not depend strongly on the adopted calibrator, agreeing in all the cases within the errors, that are considerably large in any case.

The value reported for the curvature by \citet{sanchez13} was slightly larger ($b=0.018$) than the one found in here for the same calibrator (PP04, $b=0.005$). This discrepancy is due to the limited sampling of the MZ relation at low mass in that article. They explore only 1/4 of the full CALIFA sample (150 objects, i.e., the objects observed at that date), and only 6 of them have a stellar mass lower than 10$^{9.5}$ M$_\odot$, in contrast with the current sample that contain 78 of such galaxies (a substantial number of them due to the CALIFA-entended sample). This difference highlights the importance of revisiting the MZ relation with this larger sample.
On the other hand the asymptotic metallicity  ($\rm a=8.74\pm0.01\,dex$) agrees within the errors with the currently reported one, and the $c$ parameter was chosen to match the value reported by \citet{sanchez13}, and therefore they are equal by choice.

We obtain the standard deviation along the MZ relation ($\Delta \log(\mathrm{O/H})$) for each of the calibrators by subtracting the individual metallicities measured for the 612 galaxies by the best fitted curve. This standard deviation, derived for each calibrator, is a measurement of the scatter around the estimated relation. The results of this analysis are listed in Tab.\ref{tab:val} as $\sigma$ MZ-res. These values agree with the size of the error bars shown in Fig. \ref{fig:MZ},  which indicates that the introduction of this particular functional form does not produce any significant effect in the derivation of the scatter around the mean values at each mass. As expected from the previous discussion, the abundances anchored to the direct method present a considerable lower dispersion ($\sim$ 0.05 - 0.06 dex), in comparison to those based on photoioniation models or mixed calibrators ($\sim$ 0.13- 0.18 dex). This difference in the scatter of the MZR suggests that mixed or model-based calibrators may introduce an artificial higher dynamical range of the residuals in comparison to the \temp-based ones, which nature should be explored. 

To explore the possible dependence of these results on the currently adopted functional form to describe the shape of the MZ relation, we repeated the analysis using a 4th-grade polynomial function \citep[following ][]{mann10}. We found similar standard deviations in the scatter as those reported in  Tab. \ref{tab:val}. Indeed, if we had adopted a pure spline interpolation using the mean values included in Fig. \ref{fig:MZ}, we would have found similar results. All these tests indicate that the scatter around the mean relation does not depend significantly in the adopted functional form.

\subsection{Dependence of the MZ residuals with the SFR}
\label{sec:residuals}

To explore if it is needed to introduce a secondary relation with the SFR we study the possible dependence of the residual of the MZ relation with this parameter, and if the introduction of this dependence will reduce the scatter around the new proposed relation. Any secondary relation that does not reduce the scatter is, based on the Occam's razor, not needed.

Figure \ref{fig:dMZ} shows, for each calibrator, the residual of the metallicity, once subtracted the estimated MZ relation ($\Delta\log(\mathrm{O/H})$) as a function of the SFR and the sSFR for the PPO4 calibrator. In both cases we present the median values in bins for all the analyzed calibrators. For the SFR we adopt bins of 0.3 $\log(M_{\odot} \, \mathrm{yr^{-1}})$ width in a range of -1.8 and 0.8 $\log(M_{\odot} \, \mathrm{yr^{-1}})$. For the sSFR we adopt bins of 0.4 $\log(M_{\odot} \, \mathrm{yr^{-1}})$ width in a range of -12.1 and -9.4 $\log(M_{\odot} \, \mathrm{yr^{-1}})$. The bin sizes and ranges were selected to have the same number of objects in each bin for both the SFR and the sSFR diagrams.

There is a considerable agreement in the median residuals for most of the calibrators along the considered range of SFR and sSFR, that, in all the cases are statistically compatible with a zero-value. The largest differences are found for the lowest values of SFR ($\log{\mathrm{SFR}}<-1.5$), although it does not seem to be statistically significant. For galaxies lying in the star-formation main sequence (SFMS), this SFR corresponds to stellar masses of the order of 10$^{9}$M$_\odot$ \citep[e.g.][]{mariana16}, a range where our sample is not very populated. In general there is no clear trend of the residuals with neither the SFR nor the sSFR. The only two calibrators that present a possible change with both parameters are pyqz and DOP16. However, in none of them there is a clear pattern of increasing or decreasing with the SFR, but just a fluctuation around the zero value. Moreover, we must note that these two calibrators are the ones presenting the larger error bars in the median values.

To compare with the predictions based on the two main proposed functional forms for the secondary relation with the SFR \citep{mann10,lara10a}, we overplotted them on top of our data. Following the prescriptions by \cite{mann10}, we build the blue-dashed curve by subtracting their relation without SFR dependence  (i.e., $\mu_0$ in their Eq.4) to the same relation with SFR dependence (i.e., $\mu_{0.32}$ in their Eq.4). In the same way, we build the black-dotted curve by solving the oxygen abundance in Eq.1 of \cite{lara10a}, and subtracting the MZ relation derived by the same authors (a pure second order polynomial function, described in their section 3). In both cases we took into account the differences in the IMF of the derived stellar masses, and we considered that SFR galaxies are located along the SFMS, following the functional form proposed by \citet{mariana16}. The final plots highlight the fact that the secondary relation of the SFR is more evident at low stellar masses \citep[see Fig. 1 of][]{mann10}. This is observed in our plot. There is a clear inverse relation of the SFR and the scatter in metallicity for SFR $< 1.0 M_{\odot} \, \mathrm{yr^{-1}}$ (shown for the considered calibrator, but observed in all of them). For larger SFRs the scatter is much smaller, droping from $\sim$0.08 dex to $\sim$0.04 dex. In neither of both cases we can reproduce the predicted trends, although in the case of the \citet{mann10} relation the residuals match for the range of large masses, where the authors show that the introduction of a secondary relation with the SFR produce no significant effect, in any case. 

Despite of the fact that we find no clear trend with the SFR, we attempt  to quantify the possible relation of the residuals by performing a linear fitting with the SFR for the different considered calibrators. Table \ref{tab:val} shows the results of this analysis, including the best-fitted parameters ($\alpha$ and $\beta$ for the zero point and slope, respectively), together with the standard deviation of the sample of points once removed this possible linear relation ($\sigma \Delta$MZ-res). The zero-point of the relation is compatible with zero for all the considered calibrators, and the slope for most of them. Only in the case of the pyqz calibrator we find significant positive slope, which points towards a trend oposite to the one reported in previous proposed dependences with SFR. However, when we analyze the residual after taking into account this secondary relation in all of them there is no decrease in the dispersion, even for the pyqz calibrator. Thus, the secondary relation does not produce any improvement.

\begin{figure*}
 \minipage{0.98\textwidth}
 \includegraphics[width=\linewidth]{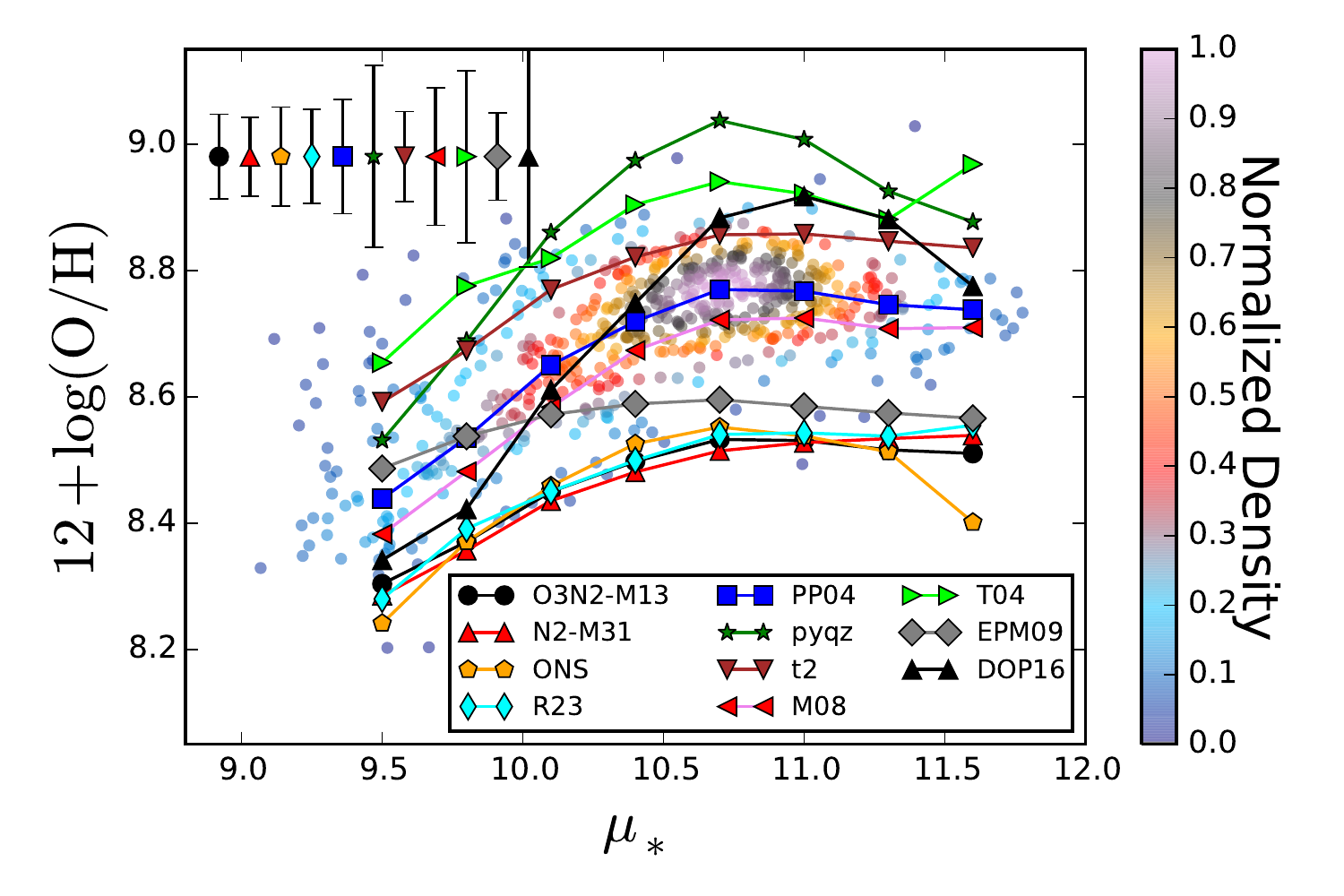}
 \endminipage
\caption{Proposed FMR relation using different metallicity calibrators for the 612 galaxies extracted from the CALIFA final sample analyzed in this study. All symbols are similar to the ones shown in Figure \ref{fig:MZ}.} 
 \label{fig:FMR}
\end{figure*}

\subsection{Exploring the FMR in detail}
\label{sec:FMR}

In the previous section we have shown that the residuals of the MZ relation derived using a set of different calibrators for the analyzed galaxies do not present any clear trend with neither the SFR nor the sSFR, at least for stellar masses larger than M$>$10$^{9.5}$M$_\odot$. However, for doing this test we adopted a linear functional form for the proposed secondary dependence with those parameters. We adopted this functional form because it is the simplest one to characterize the shape of the possible dependence. Therefore, in a purist way, our results disagree with a possible {\it linear} secondary relation with the SFR, i.e., with the functional form proposed by \citet{lara10a}.

\begin{table*}
\caption{Best fitted parameters for the $\mu_*$Z relation, adopting the numerical value for the dependence with the SFR reported by \citet{mann10} ($d=-0.32$, in Eq.\label{eq:FMR}), and the $\mu_{*,d}$Z relation, fitting the parameter $d$ (Eq.\label{eq:NEW_FMR}). For each calibrator we list the parameters $a$ and $b$ from the fitting to Eq.\ref{eq:fit} and the  $\sigma$ MZ-res, i.e., the standard deviation of the residuals after subtracting the best fit to the $\mu_*$Z relation. We include the third decimal in the $\sigma$ to highlight any possible difference. However below the 2nd decimal it is totally insignificant.}
\label{tab:FMR}
\begin{tabular} {c c c c c c c c}
\hline
Metallicity   & \multicolumn{2}{c}{$\mu_*$Z Best Fit} & $\sigma$ FMR-res & \multicolumn{3}{c}{$\mu_{*,d}$Z Best Fit} & $\sigma_d$ FMR-res  \\
 Indicator & $a$ (dex)  & $b$ (dex $/ \log(M_{\odot}))$ & (dex)  & $a$ (dex)  & $b$ (dex $/ \log(M_{\odot}))$ & $d$ (dex $/\log(M_{\odot}/yr)$ & (dex) \\
\cline{2-3}
\cline{5-7}
O3N2-M13 & 8.54 $\pm$ 0.01 & 0.011 $\pm$ 0.002 & 0.059 & 8.54 $\pm$ 0.01 & 0.016 $\pm$ 0.003 & -0.489 $\pm$ 0.108 & 0.057\\
PP04     & 8.78 $\pm$ 0.01 & 0.017 $\pm$ 0.002 & 0.085 & 8.79 $\pm$ 0.02 & 0.023 $\pm$ 0.004 & -0.489 $\pm$ 0.090 & 0.082 \\
N2-M13   & 8.54 $\pm$ 0.01 & 0.013 $\pm$ 0.001 & 0.053 & 8.54 $\pm$ 0.01 & 0.019 $\pm$ 0.003 & -0.502 $\pm$ 0.087 & 0.049\\
ONS      & 8.56 $\pm$ 0.02 & 0.015 $\pm$ 0.002 & 0.081 & 8.56 $\pm$ 0.02 & 0.016 $\pm$ 0.005 & -0.249 $\pm$ 0.246 & 0.080\\
R23      & 8.57 $\pm$ 0.01 & 0.014 $\pm$ 0.001 & 0.059  & 8.57 $\pm$ 0.02 & 0.019 $\pm$ 0.004 & -0.494 $\pm$ 0.148 & 0.057\\
pyqz     & 9.00 $\pm$ 0.04 & 0.022 $\pm$ 0.005 & 0.147  & 9.00 $\pm$ 0.02 & 0.021 $\pm$ 0.005 & -0.252 $\pm$ 0.124 & 0.146\\
t2       & 8.86 $\pm$ 0.01 & 0.013 $\pm$ 0.002 & 0.060 & 8.87 $\pm$ 0.01 & 0.020 $\pm$ 0.003 & -0.495 $\pm$ 0.090 & 0.058\\
M08      & 8.74 $\pm$ 0.01 & 0.018 $\pm$ 0.002 & 0.086 & 8.75 $\pm$ 0.02 & 0.026 $\pm$ 0.005 & -0.565 $\pm$ 0.112 & 0.085\\
T04      & 8.94 $\pm$ 0.01 & 0.014 $\pm$ 0.002 & 0.130 & 8.94 $\pm$ 0.02 & 0.013 $\pm$ 0.005 & -0.076 $\pm$ 0.226 & 0.130\\
EPM09    & 8.60 $\pm$ 0.01 & 0.005 $\pm$ 0.001 & 0.058 & 8.60 $\pm$ 0.01 & 0.007 $\pm$ 0.003 & -0.609 $\pm$ 0.254 & 0.057\\
DOP16    & 8.90 $\pm$ 0.04 & 0.030 $\pm$ 0.004 & 0.179 & 8.91 $\pm$ 0.02 & 0.040 $\pm$ 0.006 & -0.424 $\pm$ 0.072 & 0.173 \\
\hline
\end{tabular}
\end{table*}

However, due to the non linear shape of the MZ relation it is still possible that our data agree with the functional form proposed by \citet{mann10}, in particular, if it is the dispersion across the SFMS the parameter to take into account for this secondary relation as proposed by \citet{2014ApJ...797..126S}. To explore this possibility we adopted the prescriptions by \citet{mann10} and estimate the $\mu_*$ parameter for our galaxies, defined as 
\begin{equation}
\mu_* = \log(M/M_\odot) + \alpha\log(SFR)
\label{eq:FMR}
\end{equation}
with $\alpha=-0.32$, and analyze the $\mu_*$Z relation. 

Figure \ref{fig:FMR} shows the distribution of oxygen abundances along the $\mu_*$ parameter for the PP04 calibrator together with the median values of the abundances for different bins of $\mu_*$ for the remaining analyzed calibrators (with each bin having a width of $\Delta\mu_*$0.3 dex, and covering the range between $9.3-11.6$ dex in $\mu_*$). Like in the case of Fig. \ref{fig:MZ} there is a considerable agreement between the shapes of the $\mu_*$Z relation for the different considered calibrators, showing the same patters/differences already highlighted for the MZ relations described in Sec. \ref{sec:MZ}. 

We characterize the $\mu_*$Z relation using the same equation that we used to characterize the MZ relation, i.e., Eq.\ref{eq:fit}. The results of the derived asymptotic parameter ($a$) and curvature ($b$), together with the standard deviation of the residual once subtracted the best fitted curve are listed in Table \ref{tab:FMR}. When comparing with the values reported for the MZ relation (Tab. \ref{tab:val}), in general it is found that the asymptotic value does not change within the errors. On the other hand, the curvature increases for all the calibrators, like if the horizontal axis has shrunk. This is expected in general. If the SFR is correlated with the stellar mass following the SFMS, then:

\begin{equation}
\log{(SFR/M_\odot/yr^{-1})}\sim \alpha\log{(M/M_\odot)}
\label{eq:SFMS}
\end{equation}
where $\alpha$ is of the order 0.86 \citep[][]{mariana16}. In this case, at a first order $\mu_*\sim 0.72\log{M/M_\odot}$, producing a squeeze, what explains the increase of the curvature. 

More interesting is the comparison of the standard deviation of the residuals for the $\mu_*$Z relation compared with the MZ one. In general there is no significant decrease in the standard deviations. In most of the cases the values are equal, or the improvement, if any, affects the 3rd decimal. The largest difference is found for the DOP16 calibrator, with an improvement of $\Delta\sigma=$0.013 dex in the standard deviation (a reduction of $\sim$7\% in the dispersion). In summary, this analysis shows that the introduction of a secondary relation between the MZ with the SFR using the exact functional form and numerical value of the corrections described by \citet{mann10} does not produce any significant reduction of the standard deviation of the residuals over the whole distribution.

It is still possible that the adopted functional form for the FMR produces a significant improvement of the MZ relation when considering a different numerical value for the correction 
of the stellar mass in the formula. We explore that possibility by refitting the data using the following equation:

\begin{equation}
\mathrm{y}=a+b(\mathrm{x}+d\mathrm{s}-c)\exp(-(\mathrm{x}-c))
\label{eq:fit_FMR_alpha}
\end{equation}
where $y$, and $x$ are the same parameters described in Eq. \ref{eq:fit} (i.e., the oxygen abundance and the logarithm of the stellar mass), and $s$ is the logarithm of the star-formation rate $s =\log(SFR/M_{\odot}/yr)$. Like in the previous case we fix parameter $c$ to 3.5 and fit $a$, $b$ and $d$. In this formalism, parameter $d$ corresponds to the parameter $\alpha$ described by \citet{mann10}. Hereafter we refer to that parametrization as the $\mu_{*,d}$Z relation. Results are listed in Tab. \ref{tab:FMR}. In general the best fitted parameters describe a slightly stronger dependence on the SFR than the one proposed by \citet{mann10}, with values near to the one reported by those authors. The most similar values are reported for the ONS and the {\sc pyqz} calibrators, with a value of $d\sim$-0.25. However, in some cases the dependence with the SFR is much weaker (e.g., for the T04 calibrator, where $d=-$0.076). Once more, however, for none of the calibrators there is a general improvement in the standard deviation of the residuals in the overall distribution.

\begin{figure*}
 \minipage{0.995\textwidth}
 \includegraphics[width=\linewidth]{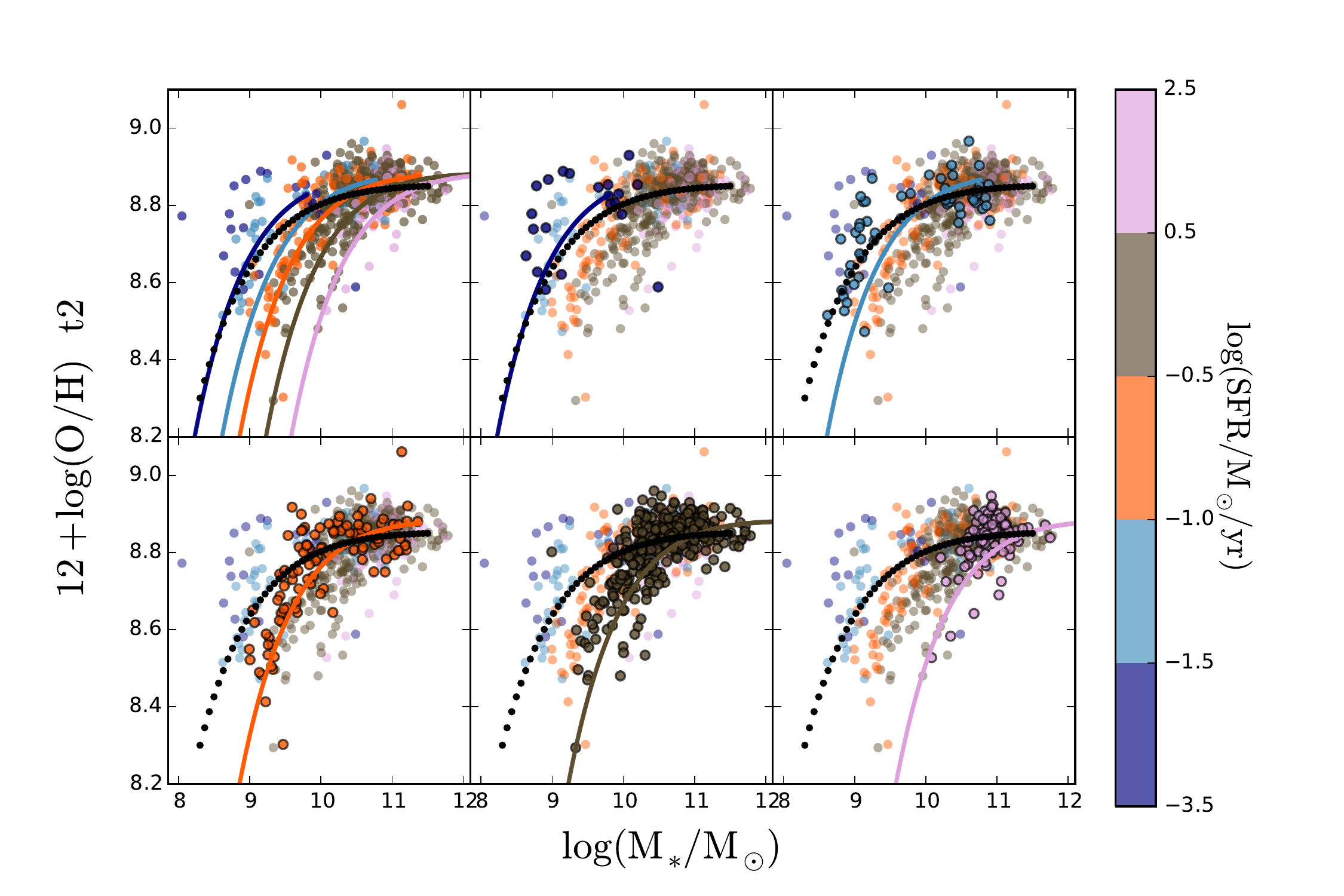}
 \endminipage
\caption{MZ distribution for the $t2$ indicator for the 612 galaxies extracted from the CALIFA final sample analyzed in this study. Each galaxy is represented by a solid circle in each panel, with colors indicating the corresponding SFR within each bin indicated in the colorbar. The first panel (top-left) shows the full distribution of galaxies together with the best fitted $\mu_{*,d}$Z relation for the average SFR within each bin, represented as a colored solid line (Eq. \ref{eq:NEW_FMR}, Tab. 2). The back dotted line represents the best fitted MZ relation, without considering any dependence with the SFR (Eq. \ref{eq:fit}, Tab. \ref{tab:val}. Each of the remaining panels shows the same distribution highlighting the galaxies and result of the fit for each of the considered SFR bins, from low star-formation rates (top-middle) to high star-formation rates (bottom-right).} 
 \label{fig:NEW_FMR}
\end{figure*}

Figure \ref{fig:NEW_FMR} illustrates the result of this analysis for one particular calibrator ($t2$). Although similar results are found for the rest of calibrators. We include only one just for clarity. We selected this particular one since it is the one that presents the smaller errors in the derived values for the pameters fitted in Eq. \ref{eq:fit} (Tab. \ref{tab:val}), being one of the three with the smaller dispersion around that relation. In other worlds, it is the one of the calibrators that fits better to the proposed functional form for the MZ-relation. Fig. \ref{fig:NEW_FMR} shows the distribution of the oxygen abundance along the stellar mass for five different bins of star-formation rates together with the location of the best-fitted $\mu_{*,d}$Z, together with the best-fitted MZ-relation (Tab. \ref{tab:val}). Despite of the results listed in Tab. \ref{tab:FMR}, by eye, it seems that including a secondary relation with the SFR using the functional form proposed by \citet{mann10} makes the model to match better with the data. The main difference between the MZ-relation and the (adjusted) FMR is found at stellar masses lower than $<10^{9.5}$M$_\odot$. However, we should advice against a conclusion guided by visual inspection to the data. Table \ref{tab:FMR_t2} shows the comparison between the residuals of the best-fitted $\mu_{*,d}$Z and MZ-relations shown in Fig. \ref{fig:NEW_FMR} for the considered star-formation bins. Contrary to what the visual inspection may indicate the residuals from the MZ-relation have a central value more consistent with the zero with a similar or smaller standard deviations around this central value for all the considered bins up to $\log(SFR)>-1.5\log(SFR/M_{\odot}/yr)$. Below that limit, i.e., the range for which we describe a possible offset in the residuals of the MZ-relation along the star-formation rate shown in Fig. \ref{fig:dMZ}, the residuals of the central value of these residuals is only 0.024 dex larger than the value derived adopting a dependence with the SFR. This value is well below the 1$\sigma$ limit ($\sigma\sim$0.12 dex). Even more, the dispersion around this central value is slightly larger for the $\mu_{*,d}$Z relation than for the MZ one. 

\begin{table}
\caption{Comparison of the residuals between the best fitted $\mu_{*,d}$Z relation (Tab. \ref{tab:FMR}) and the MZ-relation (Tab. \ref{tab:val}) for the five different star-formation bins shown in Fig. \ref{fig:NEW_FMR}. We include the range of SFRs covered by each bin, together with the average and standard deviations of the corresponding star-formation rates ($<$SFR$>$), the residuals of the $\mu_{*,d}$Z relation and the residuals of the MZ-relation. Like in previous tables we include the third decimal in the $\sigma$ to highlight any possible difference. However beyond the 2nd decimal it is totally insignificant. The same comparison was repeated for five different mass bins.}
\label{tab:FMR_t2}
\begin{tabular} {c c c c }
\hline
$\log(SFR/M_\odot/yr)$   &$<\log(SFR)>$& $\mu_{*,d}$Z-res & MZ-res \\
range & $log(M_\odot/yr)$ & (dex) & (dex) \\
\hline
$[$-3.5,-1.5$]$ & -2.01  $\pm$ 0.36  & 0.062  $\pm$ 0.126 &  0.086  $\pm$ 0.122\\
$[$-1.5,-1.0$]$ & -1.23  $\pm$ 0.15  & 0.069  $\pm$ 0.119 &  0.015  $\pm$ 0.070\\
$[$-1,-0.5$]$   & -0.73  $\pm$ 0.15  & 0.042  $\pm$ 0.082 & -0.022  $\pm$ 0.080\\
$[$-0.5,0.5$]$  &  0.02  $\pm$ 0.28  & 0.039  $\pm$ 0.056 & -0.015  $\pm$ 0.057\\
$[$0.5,0.5$]$   &  0.73  $\pm$ 0.16  & 0.047  $\pm$ 0.054 & -0.011  $\pm$ 0.047\\
\hline
$\log(M_*/M_\odot)$   &$<\log(SFR)>$& $\mu_{*,d}$Z-res & MZ-res \\
range & $log(M_\odot/yr)$ & (dex) & (dex) \\
\hline
$[$8.0,9.5$]$   & -1.11  $\pm$ 0.52  & 0.117  $\pm$ 0.110 & -0.022  $\pm$ 0.153\\
$[$9.5,10.0$]$  & -0.60  $\pm$ 0.42  & 0.058  $\pm$ 0.076 & -0.046  $\pm$ 0.095\\
$[$10.0,10.5$]$ & -0.18  $\pm$ 0.50 &  0.043  $\pm$ 0.068 & -0.023  $\pm$ 0.072\\
$[$10.5,11.0$]$  & 0.05  $\pm$ 0.56  &  0.035  $\pm$ 0.049 & 0.011  $\pm$ 0.039\\
$[$11.0,12.5$]$  & 0.12  $\pm$ 0.57  &  0.004  $\pm$ 0.048 & 0.006  $\pm$ 0.035\\
\hline
\end{tabular}
\end{table}

Finally, we repeated the comparison between the two relations shown in Fig. \ref{fig:NEW_FMR} in five mass bins. The results are listed in Tab. \ref{tab:FMR_t2}. For each mass bin we include the same parameters: the average SFR at each bin and the average and standard deviations of the residuals once subtracted each of the compared relations. The results of this comparison indicate that including a secondary relation with the SFR do not improve the quality of the MZ-relation for stellar masses higher than $>10^{10}$M$_\odot$. Indeed the standard deviation around the central value is slightly larger and the offset of this central value is of the same order or larger. For the stellar masses below this limit the dispersion around the central value decreases when taking into account the possible dependence with the SFR. However, the offset of the central value with respect to zero (i.e., the optimal value if the function is a good representation of the distribution) does not decrease significantly. Actually, for the lowest stellar masses this central value is off the zero value, being at $\sim$1$\sigma$ from zero.

\subsection{Exploring the Mass-SFR-Z plane in detail}
\label{sec:FP}

As we quoted before, \citet{lara10a} introduced the dependence of the oxygen abundance with the stellar mass and the star-formation rate using a totally different approach. Instead of modifying any MZ-relation including a term that depends on the SFR, they explored the possibility that the three parameters were distributed along a plane that they called the Mass-SFR-Oxygen abundance Fundamental Plane, with the form:

\begin{equation}
\mathrm{\log(M_*/M_\odot)}=\alpha\mathrm{\log(SFR/M_\odot/yr)}+\beta\mathrm{(12+log(O/H))}+\gamma
\label{eq:ll_FP}
\end{equation}
finding that the best fitted parameters where $\alpha=1.122$, $\beta=0.474$ and $\gamma=-0.097$. They claim that this relation presents a much smaller scatter ($\sim$0.16 dex) than the MZ relation, that has a standard deviation of 0.26 dex for their dataset. Actually, more than modifying the MZ-relation, this relation modifies the so-called star-formation main sequence (SFMS), which is a linear relation between the SFR and the stellar mass with a slope of $\sim$0.8 \citep[e.g.][and references in there]{mariana16}. By solving the SFR from the equation of \citet{lara10a} it is possible to derive a very similar slope. The novelty of this relation is that they propose that the oxygen abundance should present a linear dependence with the stellar mass once introduced the dependence with the SFR. Solving the oxygen abundance from their equation, the dependence with the other two parameters should be:

\begin{equation}
\mathrm{12+log(O/H)}=a\,\mathrm{\log(M_*/M_\odot)}-b\,\mathrm{\log(SFR/M_\odot/yr)}+c
\label{eq:fit_FP}
\end{equation}
with $a=2.110$, $b=-2.367$ and $c=0.205$. Following the same scheme of the previous section, we try to reproduce the results by \citet{lara10a} using our data. We already showed in Fig.\ref{fig:dMZ} that adopting the numerical values for their relation we cannot reproduce our observed data. However, it may still be possible that the actual values of those parameters depends on the adopted calibrator, and therefore we need to find the best fitted ones for our current dataset. For doing so, we fitted equation \ref{eq:fit_FP} to our data. 

\begin{table}
\caption{Best fitted parameters for M-SFR-Z linear relation, adopting the functional form proposed by \citet{lara10a} (Eq.\label{eq:fit_FP}). For each calibrator we list the derived parameters $a$, $b$ and $c$ of the considered equation and the standard deviation of the residuals after subtracting the best fitted model ($\sigma$ FP-res). We include the third decimal in the $\sigma$ to highlight any possible difference. However below the 2nd decimal it is totally insignificant.}
\label{tab:FP}
\begin{tabular} {c c c c c}
\hline
Metallicity   & \multicolumn{3}{c}{FP Best Fit} & $\sigma$ FP-res\\
 Indicator & $a$  & $b$  & $c$  & (dex) \\
\cline{2-4}
O3N2 & 0.10 $\pm$ 0.02 & -0.020 $\pm$ 0.019 & 8.246 $\pm$ 0.051 & 0.067 \\
PP04 & 0.14 $\pm$ 0.02 & -0.028 $\pm$ 0.022 & 8.355 $\pm$ 0.061 & 0.096 \\
N2   & 0.12 $\pm$ 0.02 & -0.039 $\pm$ 0.018 & 8.172 $\pm$ 0.050 & 0.065 \\
ONS  & 0.09 $\pm$ 0.03 &  0.003 $\pm$ 0.033 & 8.264 $\pm$ 0.081 & 0.093 \\
R23  & 0.11 $\pm$ 0.03 & -0.037 $\pm$ 0.026 & 8.217 $\pm$ 0.067 & 0.078 \\
pyqz & 0.13 $\pm$ 0.03 &  0.054 $\pm$ 0.029 & 8.587 $\pm$ 0.077 & 0.151 \\
t2   & 0.12 $\pm$ 0.02 & -0.028 $\pm$ 0.020 & 8.511 $\pm$ 0.054 & 0.077 \\
M08  & 0.14 $\pm$ 0.03 & -0.049 $\pm$ 0.025 & 8.311 $\pm$ 0.073 & 0.105 \\
T04  & 0.10 $\pm$ 0.03 &  0.002 $\pm$ 0.029 & 8.626 $\pm$ 0.083 & 0.137 \\
EPM09& 0.03 $\pm$ 0.02 & -0.008 $\pm$ 0.019 & 8.497 $\pm$ 0.052 & 0.069 \\
DP09 & 0.26 $\pm$ 0.03 & -0.025 $\pm$ 0.032 & 8.116 $\pm$ 0.086 & 0.187 \\
\hline
\end{tabular}
\end{table}

The results from this analysis are listed in Table \ref{tab:FP}, including for each calibrator the best fitted parameters and the standard deviation of the residuals once subtracted the model. The comparison of the standard deviation with those listed in Tab. \ref{tab:val}, corresponding to the MZ-relation, shows that the introduction of this functional does not produce any improvement on the modeling of the data. There is no decrease of the standard deviation for any of the considered calibrators. It is still the case that the residuals are of the same order in both cases. Moreover, by inspecting the reported parameters it is seen that most of the dependence is on the mass ($a\sim$0.1), with very little contribution of the SFR ($b\sim -$0.03). Indeed, if it is considered a simple linear dependence with the stellar mass the derived standard deviations are very similar to the reported ones. Finally, we would like to highlight that for none of the considered calibrators we can reproduce the very strong dependence of the oxygen abundance reported by \citet{lara10a} with both the stellar mass and the star-formation rate.

\subsection{Alternative exploration of the dependence of the MZR with the SFR and the sSFR}
\label{sec:salim}

\citet{2014ApJ...797..126S} proposed a different approach to explore the possible dependence of the MZ relation with the SFR in which the dependence of the Mass of both the oxygen abundance and the SFR are minimized or removed. In this particular case it is not required to assume a particular functional form for the possible dependence, like in the analysis performed in the previous section. Following these authors we start selecting only those galaxies that are located along the SFMS. For doing so we use as demarcation line the average between SFMS and the Retired Galaxies Main Sequence, proposed by \citet{mariana16}, and select only those galaxies for which:

\begin{equation}
\log{(SFR/M_\odot/yr^{-1})}>-9.58+0.835\log{(M/M_\odot)}
\label{eq:sSFR_cut}
\end{equation}

This reduces the sample to 492 pure star-forming galaxies. Then, we remove the dependence of the sSFR with the mass using the SFMS relation in \citet{mariana16}, transformed for the sSFR:

\begin{equation}
\Delta\log{(sSFR)}=\log{(sSFR/yr^{-1})}+8.34+0.19\log{(M/M_\odot)}
\label{eq:Delta_sSFR}
\end{equation}

This new parameter, the residual of the sSFR across the SFMS, does not include any dependence with the mass by construction, and therefore seems to be promising for detecting any possible correlation of the oxygen abundance with the SFR without the contamination of any relation with the stellar mass.

\begin{figure}
\includegraphics[width=\columnwidth]{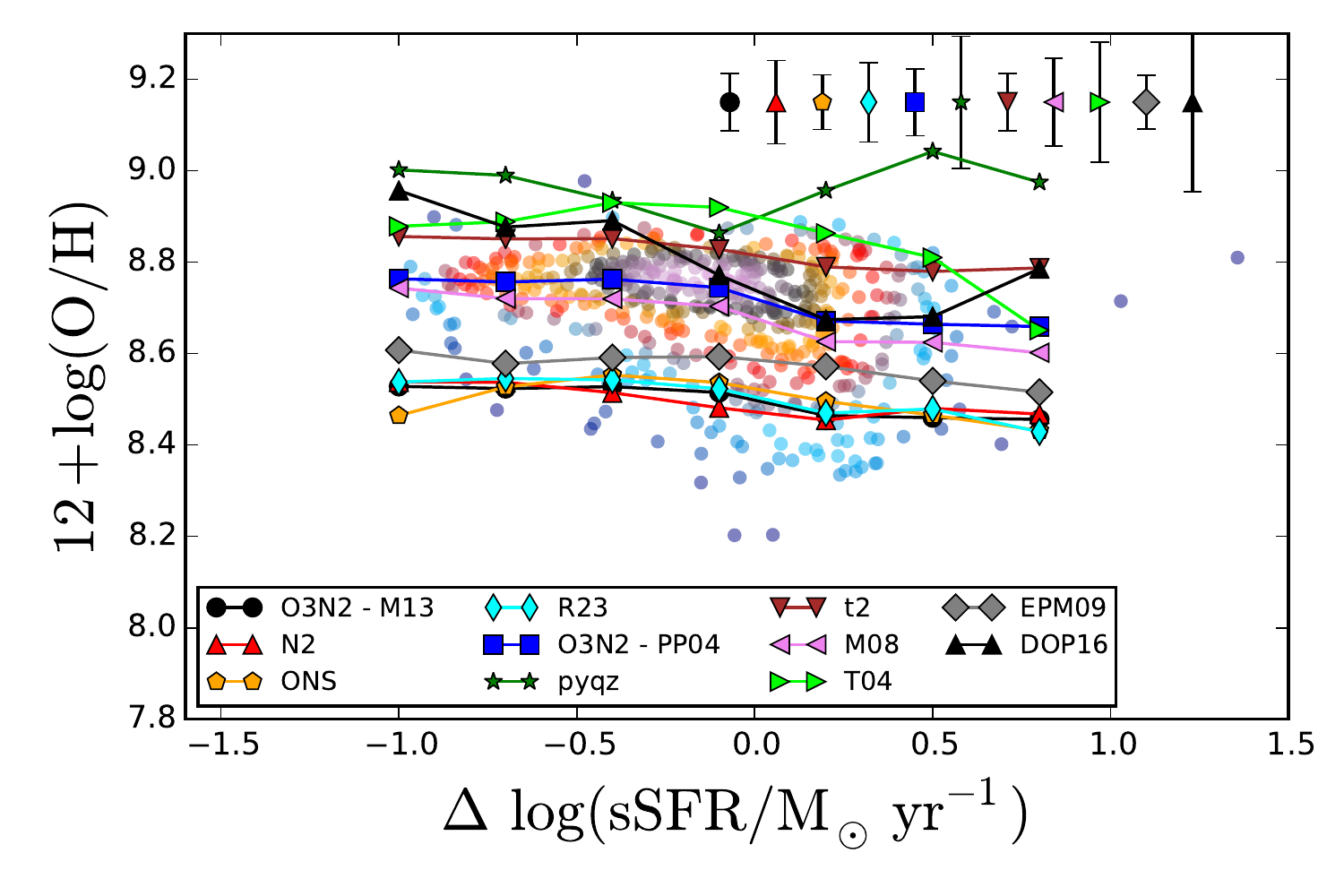}

\includegraphics[width=\columnwidth]{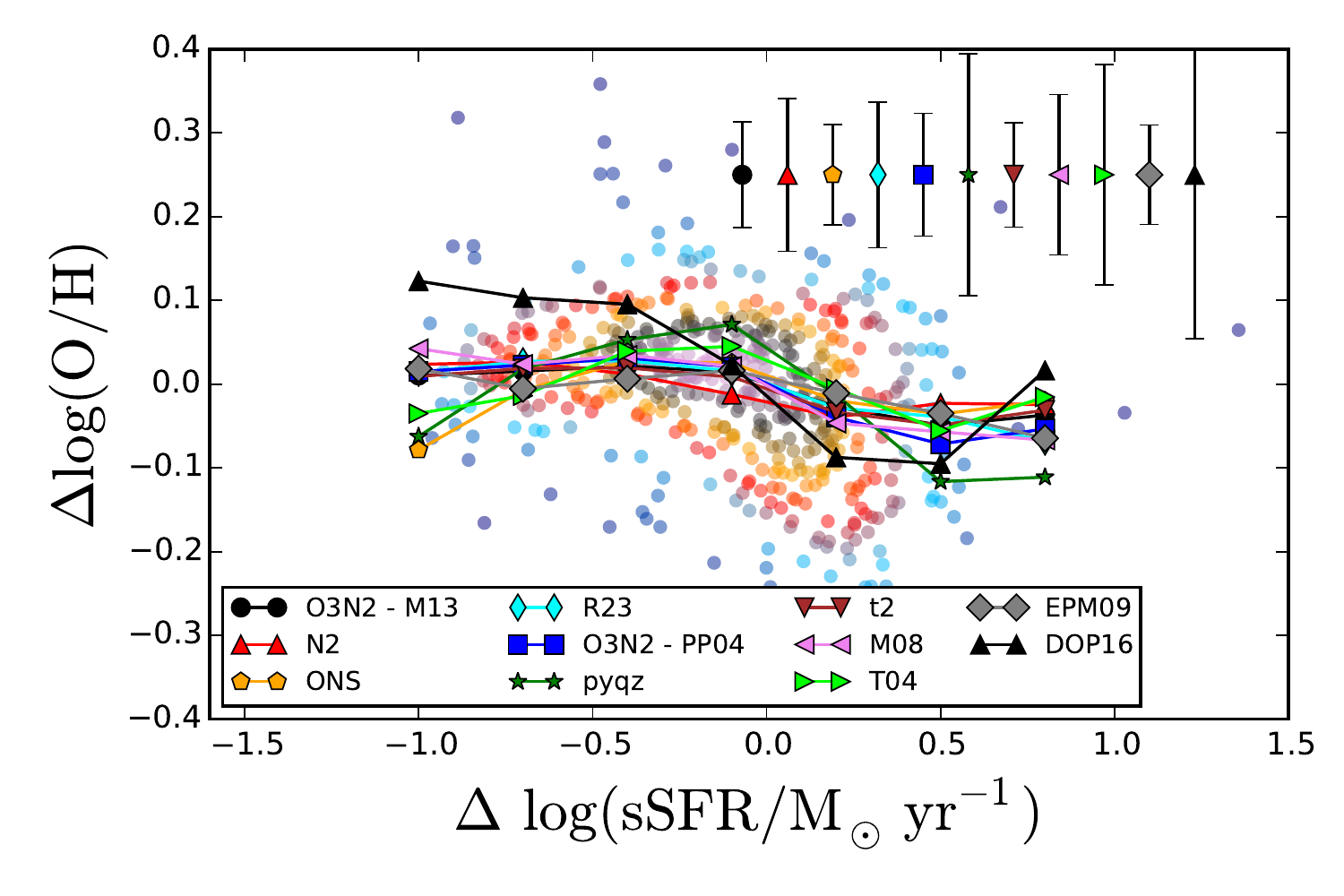}

\caption{{\it Top panel:} Distribution of the oxygen abundance along the residual of the sSFR once subtracted the dependence with the stellar mass (i.e., the SFMS), for the individual galaxies (colored dotted circles) for the PP04 calibrator together with the median oxygen abundance in bins of the residual of the sSFR (line-connected symbols) for the different calibrators. {\it Bottom panel:} Residual of the oxygen abundance once subtracted the dependence with the stellar mass (i.e., the MR relation), along the residual of the sSFR once subtracted the dependence with the stellar mass, for the individual galaxies (colored dotted circles) for the PP04 calibrator, together with the median of the residual of the oxygen abundance across the MZ relation in bins of the residual of the sSFR (line-connected symbols) for all the considered calibrators. Symbols are similar to the ones shown in Fig.\ref{fig:dMZ}. } 
\label{fig:dMZ_M}
\end{figure}

Figure \ref{fig:dMZ_M}, top panel, shows the distribution of the oxygen abundance along $\Delta\log{(sSFR)}$ parameter, including the values for individual galaxies for the PP04 calibrator and the median values in bins of 0.3 dex of the considered parameter. 
The figure shows that there are clear trends between the oxygen abundance and the  $\Delta\log{(sSFR)}$ parameter, with most of them showing an anticorrelation between both paramters. In general, all calibrators based just on R23 (e.g., R23, T04 or M08), or those that present secondary corrections due to the strength of the ionization (DOP16) show a stronger decrease of the oxygen abundance with the residuals of the sSFR, apparently supporting an FMR dependence with the SFR: the galaxies with more SFR for a certain mass would present a lower oxygen abundance. Finally, there are calibrator for which there is no clear trend (pyqz) or a weak one (ONS). Similar results were found by \cite{2014ApJ...797..126S}, who describe a different pattern of the dependence between these two parameters for different calibrators (Fig. 6 and 7 of that article). 

\begin{table}
\caption{Results of the analysis of the dependence of the oxygen abundance with the residuals of the sSFR once subtracted the dependence with the mass (i.e., the SFMS), for the different calibrators included in Fig. \ref{fig:dMZ_M}. It is listed the zero-point ($\alpha$) and the slope ($\beta$) from the linear regression, together with the standard deviation of the residuals, once subtracted the derived relation ($\sigma$-res).}
\label{tab:salim}
\begin{tabular} {c c c c}
\hline
Metallicity   & \multicolumn{2}{c}{12+$\log{O/H}$ vs. $\Delta\log{(sSFR)}$} & $\sigma$-res  \\
 Indicator & $\alpha$ (dex) & $\beta$ (dex $/ \log(yr^{-1}))$ &  (dex)\\
\cline{2-3}
O3N2-M13& 8.48 $\pm$ 0.01 & -0.04 $\pm$ 0.01 &  0.096\\
PP04    & 8.70 $\pm$ 0.01 & -0.06 $\pm$ 0.01 &  0.140\\
N2-M13  & 8.48 $\pm$ 0.01 & -0.05 $\pm$ 0.01 &  0.103\\
ONS     & 8.47 $\pm$ 0.02 & -0.02 $\pm$ 0.02 &  0.099\\
R23     & 8.49 $\pm$ 0.01 & -0.06 $\pm$ 0.01 &  0.094\\
pyqz    & 8.98 $\pm$ 0.02 &  0.01 $\pm$ 0.04 &  0.179\\
t2      & 8.81 $\pm$ 0.01 & -0.04 $\pm$ 0.01 &  0.097\\
M08     & 8.66 $\pm$ 0.01 & -0.08 $\pm$ 0.01 &  0.135\\
T04     & 8.81 $\pm$ 0.03 & -0.12 $\pm$ 0.05 &  0.180\\
EPM09   & 8.56 $\pm$ 0.01 & -0.04 $\pm$ 0.01 &  0.073\\
DOP16   & 8.77 $\pm$ 0.03 & -0.13 $\pm$ 0.05 &  0.309\\
%
%
\hline
\end{tabular}
\end{table}

We perform a linear fitting in order to quantify the trends observed in Fig. \ref{fig:dMZ_M}, top panel, and its effect in the dispersion once applied. Results of this analysis are shown in Table \ref{tab:salim}, including the zero-point and slope of the linear fitting, together with the standard deviations of the distributions after applying the estimated relation. As indicated before, for most of the calibrators there seems to be an anticorrelation between the two parameters. However, when we analyze the effects in the dispersion (by comparing with the original standard deviation listed in Tab.\ref{tab:val}), the introduction of this relation does not improve the dispersion for any calibrator. 

The previous analysis is not a test in favor or against any possible secondary relation between the MZ with the SFR or the sSFR. Indeed, it explores a possible primary relation between the oxygen abundance and the dispersion of the sSFR along the SFMS. As indicated before, our results, and that of \citet{2014ApJ...797..126S}, show that if there is such a relation, it is less general than the MZ relation, since it depends on the adopted calibrator (not only on the shape, but on the global trend), and it does not decrease the original dispersion as much as the MZ relation itself (compare $\sigma$ MZ-res parameter from Tab.\ref{tab:val} with $\sigma$-res parameter from Tab. \ref{tab:salim}).

To explore if there is a relation of the oxygen abundance with the SFR totally independent of the mass it is needed to remove that dependence in the two analyzed parameters. For doing so \citet{2014ApJ...797..126S} separated the analyzed sample in bins of stellar masses of $\sim$0.5 dex and performed the same test. When doing so, adopting the O3N2-PP04 on the published CALIFA data up to that date \citep{sanchez13}, they found a negative trend in three bins and a positive trend in the remaining one (fig. 10 of that article). However, for the full mass range of the adopted SDSS sub-sample they found a positive trend for all mass bins. 

\begin{table}
\caption{Results of the analysis of the dependence of the oxygen abundance residuals with respecto to the MZ relation along the residuals of the sSFR once subtracted the dependence with the mass (i.e., the SFMS), for the different calibrators included in Fig. \ref{fig:dMZ_M}. It is listed the zero-point ($\alpha$) and the slope ($\beta$) from the linear regression, together with the standard deviation of the original distribution ($\sigma$-org) and that of the residual, once subtracted the derived relation ($\sigma$-res).}
\label{tab:salim2}
\begin{tabular} {c c c c}
\hline
Metallicity   & \multicolumn{2}{c}{$\Delta\log{O/H}$ vs. $\Delta\log{(sSFR)}$} & $\sigma$-res  \\
 Indicator & $\alpha$ (dex) & $\beta$ (dex $/ \log(yr^{-1}))$ & (dex) \\
\cline{2-3}
O3N2-M13 & -0.02 $\pm$ 0.01 & -0.04 $\pm$ 0.01 & 0.064\\
PP04     & -0.02 $\pm$ 0.01 & -0.05 $\pm$ 0.02 & 0.093\\
N2-M13   & -0.02 $\pm$ 0.01 & -0.04 $\pm$ 0.01 & 0.057\\
ONS      & -0.01 $\pm$ 0.02 &  0.04 $\pm$ 0.03 & 0.085\\
R23      & -0.01 $\pm$ 0.01 & -0.05 $\pm$ 0.01 & 0.063\\
pyqz     & -0.04 $\pm$ 0.03 & -0.05 $\pm$ 0.05 & 0.153\\
t2       & -0.02 $\pm$ 0.01 & -0.03 $\pm$ 0.01 & 0.064\\
M08      & -0.02 $\pm$ 0.01 & -0.07 $\pm$ 0.02 & 0.095\\
T04      & -0.01 $\pm$ 0.02 & -0.02 $\pm$ 0.03 & 0.123\\
EPM09    & -0.02 $\pm$ 0.01 & -0.03 $\pm$ 0.01 & 0.053\\
DOP16    & -0.03 $\pm$ 0.02 & -0.15 $\pm$ 0.04 & 0.184\\
%
\hline
\end{tabular}
\end{table}

In any case, taking a limited range of stellar masses may reduce the dependence on this parameter, but not remove it completely. Here we effectively remove that contribution by analyzing the possible relations between the residuals of the oxygen abundance once subtracted the MZ relation found for each calibrator ($\Delta\log{O/H}$, shown in Fig.\ref{fig:dMZ}) against the residuals of the sSFR once removed the dependence with the stellar mass ($\Delta\log{sSFR}$), for the different calibrators included in this article.
Figure \ref{fig:dMZ_M}, bottom panel, shows this distribution, including the individual points for the O3N2-PP04 calibrator together with the average points for each calibrator in the same bins of $\Delta\log{sSFR}$ included in the top panel. In this case there is no general trend for all the calibrators. For some calibrators, like R23, {\sc t2}, M08 and EPM09 there is a weak decrease, that it is stronger in the case of DOP16. However, there are calibrators without a clear trend, showing first a rising and then a decrease, like T04, ONS or pyqz. In general, the trends do not seem to have a pattern with the nature of the considered calibrators.

To quantify the dependence between the two parameters and if it has an effect on the dispersion we repeated the analysis described before, performing a linear fitting to the parameters shown in Fig. \ref{fig:dMZ_M}, bottom panel. Table \ref{tab:salim2}  shows the results of this analysis, including the zero-point and slope of the linear fitting, together with the standard deviations of the distributions before and after applying the estimated relation. The different trends described before are now quantified as positive slopes (e.g. ONS), negative ones (e.g. DOP09), or slopes compatible with zero (e.g., T04). More interesting is the comparison between the standard deviations listed in this table with the ones of the original MZ relation, listed in Tab. \ref{tab:val}. In general there is little change in the dispersion for any of the considered calibrators, and, like in the case of the trend, there is no general pattern. In three cases there is a slightly decrease, in all the cases lower than 0.01 dex, and in a similar number of cases there is an increase of the dispersion. Thus, to include a possible secondary correlation with 
$\Delta\log{sSFR}$ does not produce any significant general decrease of the dispersion of the MZ relation.

\begin{figure*}
 \minipage{0.995\textwidth}
 \includegraphics[width=\linewidth]{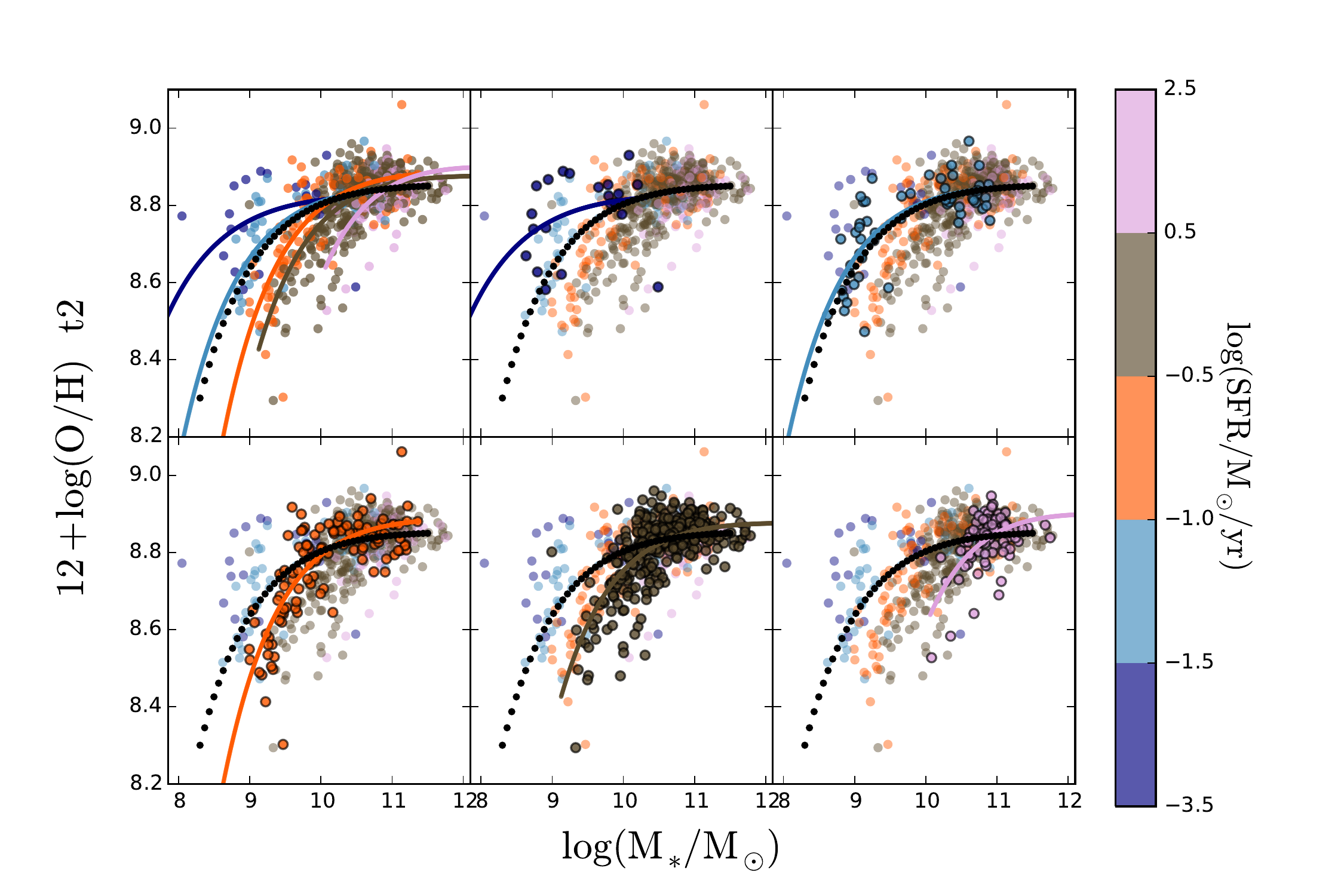}
 \endminipage
\caption{MZ distribution for the $t2$ indicator for the 612 galaxies extracted from the CALIFA final sample analyzed in this study. Each galaxy is represented by a solid circle in each panel, with colors indicating the corresponding SFR within each bin indicated in the colorbar. The first panel (top-left) shows the full distribution of galaxies together with the best fitted MZ-relation derived for each bin, represented as a colored solid line (Eq. \ref{eq:fit}, Tab. \ref{tab:MZSFR}). The back dotted line represents the best fitted MZ relation derived for the full dataset, without considering any bin in SFR
(Tab. \ref{tab:val}). Each of the remaining panels shows the same distributions highlighting the galaxies and results of the fit for each of the considered SFR bins, from low star-formation rates (top-middle) to high star-formation rates (bottom-right). For consistency we adopted the same bins than those shown in Fig. \ref{fig:NEW_FMR}.} 
 \label{fig:MZSFR}
\end{figure*}

\subsection{Dependence of the MZR with the SFR}
\label{sec:MZSFR}

In previous sections we have explored if the residuals of the MZ-relation present an evident correlation with the SFR (Sec. \ref{sec:residuals}), or with the residuals of the SFMS (Sec. \ref{sec:salim}). We also explored the possible improvements on the relation when introducing two different proposed correlations with the SFR (Sec. \ref{sec:FMR} and \ref{sec:FP}). In this section we actually explore if the parameters of our proposed functional form for the MZ-relation (Eq. \ref{eq:fit}) present a dependence with the SFR and if taking into account this dependence improves the relation in a significant way. For doing so we use the same data-set used in Sec. \ref{sec:FMR} and shown in Fig. \ref{fig:NEW_FMR}, i.e., the dataset comprising the oxygen abundances derived using the $t2$ calibrator split in five different SFR bins ($\log(SFR/M_\odot/yr)\subset$[$\infty$,-1.5],[-1.5,-1],[-1,-0.5],[-0.5,0.5],[0.5,$\infty$]). Then, for each subset we derive the best fitted parameters for Eq. \ref{eq:fit}, to see if they present any variation with the SFR. We consider that this approach is more general than assuming a particular functional form for the dependence with the SFR. If the parameters change significantly for each SFR bin, then the shape or the scale of the MZR depends on the star-formation rate. However, if the parameters do not present a statistically significant change, then the average MZR is a good description of the data, without requiring a secondary dependence with the SFR.

\begin{table}
\caption{Results of the analysis of the MZ relation for the $t2$ calibrator for the five different star-formation bins shown in Fig. \ref{fig:MZSFR}, already presented in Tab \ref{tab:FMR_t2}. We include the range of SFRs covered by each bin, the best fitted $a$ and $b$ parameters (Eq. \ref{eq:fit}) together with the average and standard deviation of the residuals of the MZ relation. The values for the full range of SFRs have been transcribed from Tab. \ref{tab:val} to facilitate the comparison. Like in previous tables we include the third decimal in the $\sigma$ to highlight any possible difference. However beyond the 2nd decimal it is totally insignificant.}
\label{tab:MZSFR_t2}
\begin{tabular} {c c c c }
\hline
$\log(SFR/M_\odot/yr)$ & \multicolumn{2}{c}{MZ Best Fit} & MZ-res \\
range & $a$ (dex)  & $b$ (dex $/ \log(M_{\odot}))$ & (dex) \\
\cline{2-3}
$[$-3.5,-1.5$]$ & 8.83 $\pm$ 0.12 & 0.002 $\pm$ 0.004 &  0.003  $\pm$ 0.105\\
$[$-1.5,-1.0$]$ & 8.85 $\pm$ 0.05 & 0.006 $\pm$ 0.003 & -0.002  $\pm$ 0.069\\
$[$-1,-0.5$]$   & 8.88 $\pm$ 0.04 & 0.013 $\pm$ 0.003 &  0.003  $\pm$ 0.070\\
$[$-0.5,0.5$]$  & 8.87 $\pm$ 0.02 & 0.017 $\pm$ 0.003 &  0.001  $\pm$ 0.049\\
$[$0.5,0.5$]$   & 8.88 $\pm$ 0.04 & 0.041 $\pm$ 0.025 &  0.001  $\pm$ 0.051\\
\hline
 any SFR        & 8.86 $\pm$ 0.01 & 0.013 $\pm$ 0.002 & -0.016  $\pm$ 0.060\\
\hline
\end{tabular}
\end{table}

The results from this analysis are shown in Figure \ref{fig:MZSFR} where the best fitted MZ-relations for the different bins of SFR are shown following the same scheme as in Fig. \ref{fig:NEW_FMR}. The derived parameters are listed in Table \ref{tab:MZSFR_t2}, including the central value (mean of the distribution) and standard deviation of the residuals for the best fitted relation for each SFR range, together with the corresponding values for the full $t2$ dataset. For a good fitting the mean or central value should be near to cero, and the standard deviation should be of the order of the typical error for the oxygen abundance ($\sim$0.3-0.05 dex). Both  parameters for the residuals should be compared with the ones listed in the last column of Tab. \ref{tab:FMR_t2}, that corresponds to the best fitted MZ-relation for the full range of SFRs. For any SFR bin there is an improvement of the fitting when using the ad-hoc MZ-relation for each subset in terms of the central value of the residual, that is closer to zero. This is particularly true for the low SFR range ($\log(SFR/M\odot/yr)<-$1.5). There is also a slightly improvement in the standard deviation, again stronger for the low star-formation values. However, we should stress than none of those improvements are statistically significant, and only in the case of the low SFR range the improvement in the central value is near to $\sim$1$\sigma$. 

Regarding the parameters describing the MZ relation, i.e., the asymptotic oxygen abundance ($a$) and the slope of the linear regime ($b$), this analysis indicates that both of them present a weak rising with the SFR. Those trends are similar to the ones described by \citet{mann10}, but including a variation in the asymptotic oxygen abundance that the parametrization presented by those authors did not take into account. Actually, the described trends indicate that the behavior of the oxygen abundances for low and high stellar masses is different for weak or strong star-formation rates. While at high stellar masses (M$>10^{10}$M$_\odot$) galaxies with low star-formation rate present lower oxygen abundances, at low stellar masses (M$<10^{10}$M$_\odot$) the trend is the opposite \citep[being similar for this range to the one described by][]{maio08}. However, like in the case of the residuals, we should stress that none of those variations seems to be statistically significant. The largest difference is found between the slope of the linear regime for the lowest and highest SFR ranges, being just $\sim$1.5$\sigma$ significant. If we take into account the number of galaxies in each bin ($\sim$100), there is a significant difference between the slopes, but they will be at less than 2$\sigma$ considering the dispersion. Like in the case of the previous tests described in previous sections there is no statistically significant improvement in the description of the data (i.e., standard deviation of the residuals) when including a possible general variation of the parameters of the adopted MZ-relation with the star-formation rate.

\section{Discussion}
\label{sec:con}

We revisit the mass-metallicity relation based on CALIFA data already explored by \citet{sanchez13}, increasing by a factor four the number of objects (734 objects). These data allow us to derive in a consistent way both the stellar mass and the characteristic metallicity at the effective radius for 612 galaxies. We adopt eleven different abundance calibrators of very different nature to determine in the most general way the shape of the MZ relation and its possible dependence with the SFR. We confirm the reported trend of the MZ relation already shown in many different publications based on both single-aperture spectroscopic surveys \citep[e.g.][]{tremonti04} and IFS ones \citep[e.g.,][]{2012ApJ...756L..31R, sanchez14b, 2016MNRAS.463.2513B}. Contrary to previous claims \citep[e.g.][]{2008ApJ...681.1183K}, we find that the MZ relation is well represented by the same functional form (i.e., the same shape) irrespectively of the adopted calibrator. The main difference is the value of the asymptotic oxygen abundance at high masses, and in the dispersion around the reported relation. This relation is considerably tighter than the one reported using single-aperture spectroscopy \citep[$\sim$0.1 dex][]{tremonti04}, having a dispersion of $\sim$0.05 dex in the tighest case.
This dispersion is of the order of the expected errors for the estimated abundances. 

In general, those calibrators anchored to the Direct Method present a tighter correlation with the stellar mass than those ones based on photoionization models. Actually, considering the adopted functional form for the MZ-relation, and taking into account that our precision in the derivation of the stellar masses is of the order of $\sim$0.1 dex, the dispersion around this relation is dominated by the errors in the stellar mass, being the tighest possible even if the oxygen abundance presents no error. If the stellar mass, the integral of the star-formation history over cosmic time, is the main driver of the current oxygen abundance in a galaxy, this difference may indicate that indeed photoionization models predict a more imprecise abundance, although the reported value may be more accurate. On the other hand, the calibrators anchored to the Direct Method may produce a more precise value for the oxygen abundance, although its value is more inaccurate. The introduction of a $t2$ correction could be a compromise solution that would make abundances based on the Direct Method more precise and accurate than the ones derived using photoionization models.

We explore different proposed scenarios for a possible secondary relation of the oxygen abundance with the star-formation rate, once considered the main relation of both parameters with the stellar mass. Among them we study: (i) the possible dependence of the residuals of the MZ relation with the SFR, (ii) the effects in the dispersion when imposing one of the most frequently adopted secondary relations \citep[the one proposed by][]{maio08}, and (iii) the possible relation of the oxygen abundance and the MZ residuals with the specific star-formation rate once removed its dependence with the stellar mass, following \citet{2014ApJ...797..126S}. In none of these cases we find a clear effect (decrease) in the dispersion of the MZ relation for the explored calibrators. We should stress that our sample is not complete/statistically significant for stellar masses below $<10^{9.5}$M$_\odot$.

In addition we explore in detail the relations between the stellar mass, SFR and oxygen abundances proposed by \citet{maio08} and \citet{lara10a}. In the first case we find that the parametrization of the possible dependence of the MZ-relation with the SFR does produce only a marginal improvement in the accuracy and precision of the relation to describe the data. This improvement, statistically not significant, it is larger for the lowest SFR and stellar masses ranges. On the other hand the parametrization proposed by \citet{lara10a} does not seem to present any improvement over the proposed MZ-relation. Finally, we explore the possible dependence of the parameters describing our proposed MZ-relation with the SFR, in the most general way. Like in the case of the relation proposed by \citet{maio08} we find a weak trend with the SFR. However the behavior seems to be different for low and high stellar masses. While at low stellar masses galaxies with low SFR seem to have larger oxygen abundances, at high stellar masses they seem to be less metal rich than the average. However, none of those trends seems to be statistically significant and they do not produce a statistically significant improvement of the description of the data compared to a MZ-relation independent of the SFR.

These results are totally consistent with the previously reported ones
by Barrera-Ballesteros et al. (submitted), in which they explored the
MZ relation for $\sim$1700 galaxies extracted from the MaNGA IFS
dataset.  Our results agree both qualitatively and quantitatively,
supporting the previous claim by \citet{sanchez13} of the lack of a
statistically significant secondary relation of the MZ with the SFR
for galaxies with stellar masses above $>10^{9.5}$M$_\odot$. We should
stress that even using single-aperture spectroscopic data this
secondary relation is under discussion. Recently, \citet{kash16}
reported that they could not find the proposed secondary relation with
the SFR, and \citet{telf16} indicated that if it exists it is 
times weaker than what it was previously claimed. Following
\citet{kash16} (in the abstract) we could claim that the fact that we
do not find the secondary relation it is not an evidence that it does
not exist. Or we could try to understand why in some cases it appears
and in others it does not.

The more evident reasons why IFS data consistently fail to reproduce
the secondary relation with the SFR of the MZ relation while
single-aperture spectroscopic data find it (at different degrees)
could be: (i) differences in the observational approaches, (ii)
differences in the observed samples, (iii) differences in the
estimations of the physical quantities (stellar mass, oxygen
abundances and star-formation rates). The first case was explored by
\citet{sanchez13} in the appendix of that article, in which the
effects of selecting a single-aperture was reproduced using the CALIFA
IFS data. A secondary relation of the same intensity as the one
reported by \citet{maio08} appears in the data when the CALIFA
galaxies are randomly shifted to the SDSS redshift range and
single-aperture spectra of 3$\arcsec$/diameter are extracted from the
cubes. Therefore, an aperture effect could be one of the primary
causes of the discrepancy. More recently \citet{telf16} found that
there is a significant aperture effect that could produce a secondary
relation of the oxygen abundance with the SFR (Section 3.3 and
Fig. 11). However, they discarded that effect without a major
reason. In any case, as pointed out, their reported secondary relation
is weaker than the one proposed before.

Differences in the observed samples could be a possible source of
discrepancy. In the current article and in the recently presented by
Barrera-Ballesteros et al. (2016) the number of galaxies is
considerably smaller than the one presented in any SDSS based
analysis. That could be a potential source of difference if the
samples have different properties. However, as shown in
Fig. \ref{fig:sample}, in \citet{sanchez13}, \citet{walcher14} and
\citet{DR3} the CALIFA sample mimic most of the properties and
distributions of the SDSS one, without major biases (for
$M_*>10{9.5}$M$_\odot$)

Another reason why there could be different results by different studies is narrowed down in the current analysis (and the one performed by Barrera-Ballesteros et al. submitted): The use of different abundance calibrators. Although our collection of calibrators is far from being complete \citep[look at ][for a different collection]{2008MNRAS.388.1321P}, we have included the most frequently used ones, covering calibrators with very different natures. In none of them we have found a clear secondary relation with the SFR. Therefore, to adopt a different calibrator does not seem to be the source of the discrepancy.

The other main reason is the differences in the adopted analysis. In some cases it was explored the decrease in the dispersion once applied the proposed secondary relation \citep[e.g.][]{maio08}, or explored the possible systematic effects in the data in detail \citep[e.g.][]{telf16}. Finally, in other cases it was proposed a secondary relation without exploring the effects in the dispersion \citep[][]{2014ApJ...797..126S}.
Those differences in the methodology may alter the interpretation. In here we describe negative and positive correlations between the residuals of the MZ relation with the SFR that in the approach of \citet{2014ApJ...797..126S} may imply a secondary correlation. However, if that correlation does not decrease the initial dispersion, from our perspective it is not required to be introduced.

In any case, the main difference compared with previous studies is the way the way that the physical parameters are derived. The SFR and the stellar masses are cumulative properties, and therefore they rely on aperture effects as indicated before. In the case of the SFR it is normally adopted a calibrator based on H$\alpha$, corrected by dust attenuation derived by the H$\alpha$/H$\beta$ ratio, assuming a certain extinction law \citep[e.g.][]{kennicutt89,catalan15}. However, for single-aperture spectroscopic data it is not possible to disentangle the different ionization sources that may contribute to the H$\alpha$ flux \citep{catalan15}. This effect is important in low resolution IFS data, as pointed out by \citet{mast14}, and recently revisited for the MaNGA dataset by \citet{zhang16}. The central region of galaxies may present many different ionization sources and even if the integrated spectra is dominated by star-formation, the contribution of other sources, like post-AGBs, shocks or AGNs \citep[e.g][]{binn94,binn09,sarzi10,sign13,belf16a,davies16} may be significant and alter the estimation of the SFR. However, recent results indicate that indeed that least the presence of an AGN do not alter significantly neither the SFR nor the oxygen abundance and its gradient \citep[e.g.][]{catalan15,lluis16}.

The stellar mass is another source of uncertainties. In many cases,
for the SDSS, it is used a stellar mass derived using multiband
photometry or the combination of that photometry with the M/L derived
from the spectroscopic information. Since the spectroscopic
information is biased towards the central regions of the galaxies
(although the fraction of light is in average $\sim$30\% at
$z\sim$0.1, at the peak of the SDSS distribution) the M/L may not be
representative of the full optical extension. In most of the cases the
uncertainties of the stellar mass are similar or larger than the ones
reported for the MZ relation \citep[typically $\sim$0.1 dex,
  e.g.][]{sanchez13,rosa14,Pipe3D_II}. We consider that the stellar
mass derived using IFS should be more accurate and precise, since the
M/L is spatially sampled and not averaged, weighting better the low
surface brightness disk regions than single aperture spectroscopic
(that weights more the bulge of galaxies).

Finally for the derivation of the oxygen abundance the mix of
ionization sources could be an important effect in the increase of the
dispersion \citep[e.g.][]{zhang16}. This is particularly important if
we take into account that the abundance is not an integrated property
of galaxies, but a relative one that presents clear gradients within
the FoV \citep[e.g.][]{sanchez14}. Even more if the gradient presents
different behaviors at different stellar masses and spatial ranges
\citep[][Belfiore et al., 2016, in prep.]{2016A&A...587A..70S}. None
of these effects can be easily removed by aperture corrections, like
the ones proposed by \citet{2016ApJ...826...71I}, and certainly they
have not been taken into account in single aperture spectroscopic
studies so far.

In the current article we adopted as a characteristic abundance for
the galaxies the value at the effective radius (that in general is
$\sim$0.05-0.1 dex lower than the central value). In single-aperture
spectroscopic surveys the abundance is dominated by the central
values, that, in addition to the possible contamination between
different ionization sources, it is not representative of the average
oxygen abundance, for the reasons indicated before \citep[mostly due
  to the drop in the oxygen abundance described for certain
  galaxies,][]{2016A&A...587A..70S}. Therefore, it could still be
possible that both results are correct at the same time, and the
presence or lack of secondary relation with the SFR is an effect of
the region of the galaxy where the abundance is derived. If this is
the case the secondary relation with the SFR found in single-aperture
spectroscopic surveys may reflect a local effect in the center of the
galaxies that does not affect the overall evolution of the
disk. Indeed, this was discussed in the the appendix of
\citet{sanchez13} as detailed before.


The most widely accepted explanation for the secondary relation with
the SFR is the presence of strong outflows connected with the
star-formation itself that spell metal-rich gas from the galaxy
decreasing the oxygen abundance{, or reaching an equilibrium
  abundance \citep[e.g.][]{belf16a}.  The lack of a secondary relation
  with the SFR open new possible interpretations.  One of them is
  that the metal enrichment is dominated by local processes, }
following the local star-formation history, with a mass-metallicity
relation driven by the local downsizing process
\citep[e.g.][]{perez13,ibarra16}. The almost linear regime of the MZ
relation (M$_*<$10$^{10}$M$_\odot$), and its local version, the
$\Sigma$-z relation \citep[$\Sigma_*<$2.5M$_\odot$pc$^{-2}$,
][]{rosales12}, may reflect the regime for which the actual sSFR is
stronger, where metal enrichment is still happening. On the other
hand, the plateu at larger stellar masses (and mass densities) is
conformed by those galaxies (or regions of galaxies) which bulk of
stellar mass was formed in early cosmological times, and thus the
enrichment (e.g., more early type spirals or the central regions of
disk galaxies). With a star-formation regulated mostly by the
reservoir of molecular gas and the inflow \citep[e.g.][]{lill13}, both
the abundance gradients and the local $\Sigma$-z relation are easily
reproduced \citep[e.g.][]{2016MNRAS.463.2513B}. 

{ However, this is
  not the only possible explanation for the shape of the
  MZ-relation. It is possible that the MZ relation has a plateu 
  becuse it reaches the maximum yield of oxygen abundance, for a
  characteristic depletion time \citep{pily07}. Even more, we still
  need to accommodate the possible secondary correlations described
  for the atomic and maybe the molecular cold gas in this picture
  \citep[][]{2012ApJ...745...66M,bothwell16}, if they are confirmed.
A plausible explanation of why there could be a secondary relation
with the molecular gas and not with the SFR would be the existence 
of different depletion times for galaxies at different stellar mass
and morphology. It is known that the KS-law \citep[e.g.][]{kennicutt98} present
a considerable scatter that it is related with that parameter, and
on-going studies indicate that the depletion time may be different
for different galaxies and in different regions of the same galaxy (Dyas et al., in prep). }

{ Finally}, it may still be possible that outflows affect the chemical
evolution of the galaxies more in the central regions than in the
outer parts. Like in the case of many other processes outflows are
local processes regulated by the local star-formation rate (and thus
the local SN rate), the local gas density, and the local values of the
escape velocity and angular momentum. For low-mass galaxies with high
star-formation it may be possible to form strong gas outflows
\citep[e.g.][]{carlos16}, that affect the oxygen abundance only in the
central regions. Under this scenario it is still possible that a
secondary relation with the SFR is found for the MZ-relation for
central abundances, and a lack of this secondary relation for the
characteristic value. We will explore that possibility in further
analyses of the data.

\section{Conclusions}

In summary, we cannot confirm the existence of an statistically
significant secondary dependence of the MZ-relation with the SFR based
on our analysis of the characteristic abundance for a sample of IFS
observed galaxies extracted from the CALIFA sample. If there is such
secondary relation it is less significant than what it was reported in
previous studies, maybe confined to the low stellar mass range
(M$<10^{9.5}$M$_\odot$), and it does not produce a significant
improvement over the MZ-relation beyond that mass range. This result
agrees with an evolution of the stellar population in galaxies and in
particular the metal enrichment dominated by local processes, mostly
regulated by the local reservoir of gas, with a limited influence of
gas outflows. If the secondary relation is confined to the central
oxygen abundances, it may indicate that the effects of outflows or the
stopping of the star-formation in the central regions modifies the
chemical enrichment in those areas in a different way than the general
evolution of the disks. { However, other possible scenarios are still
possible, in particular if the proposed secondary correlations with the
atomic and molecular gas are confirmed.}

\section*{Acknowledgements}

{ We thanks the anonymous referee for his/her comments that has helped to 
improve the manuscript }

SFS and CLC thank the ConaCyt programs IA-180125 and DGAPA IA100815
and IA101217 for their support to this project. LSM thanks support
from the Spanish {\em Ministerio de Econom\'ia y Competitividad
  (MINECO)} via grant AYA2012-31935.  L.G. was supported in part by
the US National Science Foundation under Grant AST-1311862.  RAM
acknowledges support by the Swiss National Science Foundation.
S.Z. has been supported by the EU Marie Curie Career Integration Grant
{\it SteMaGE} Nr. PCIG12-GA-2012-326466 (Call Identifier: FP7-PEOPLE-2012
CIG)

This study  uses data provided by the Calar Alto Legacy
Integral Field Area (CALIFA) survey (http://califa.caha.es/).

CALIFA is the first legacy survey performed at Calar Alto. The
CALIFA collaboration would like to thank the IAA-CSIC and MPIA-MPG as
major partners of the observatory, and CAHA itself, for the unique
access to telescope time and support in manpower and infrastructures.
The CALIFA collaboration also thanks the CAHA staff for the dedication
to this project.

Based on observations collected at the Centro Astron\'omico Hispano
Alem\'an (CAHA) at Calar Alto, operated jointly by the
Max-Planck-Institut f\"ur Astronomie and the Instituto de Astrof\'\i sica de
Andaluc\'\i a  (CSIC).

Funding for the Sloan Digital Sky Survey IV has been provided by
the Alfred P. Sloan Foundation, the U.S. Department of Energy Office of
Science, and the Participating Institutions. SDSS-IV acknowledges
support and resources from the Center for High-Performance Computing at
the University of Utah. The SDSS web site is www.sdss.org.

SDSS-IV is managed by the Astrophysical Research Consortium for the 
Participating Institutions of the SDSS Collaboration including the 
Brazilian Participation Group, the Carnegie Institution for Science, 
Carnegie Mellon University, the Chilean Participation Group, the French Participation Group, Harvard-Smithsonian Center for Astrophysics, 
Instituto de Astrof\'isica de Canarias, The Johns Hopkins University, 
Kavli Institute for the Physics and Mathematics of the Universe (IPMU) / 
University of Tokyo, Lawrence Berkeley National Laboratory, 
Leibniz Institut f\"ur Astrophysik Potsdam (AIP),  
Max-Planck-Institut f\"ur Astronomie (MPIA Heidelberg), 
Max-Planck-Institut f\"ur Astrophysik (MPA Garching), 
Max-Planck-Institut f\"ur Extraterrestrische Physik (MPE), 
National Astronomical Observatories of China, New Mexico State University, 
New York University, University of Notre Dame, 
Observat\'ario Nacional / MCTI, The Ohio State University, 
Pennsylvania State University, Shanghai Astronomical Observatory, 
United Kingdom Participation Group,
Universidad Nacional Aut\'onoma de M\'exico, University of Arizona, 
University of Colorado Boulder, University of Oxford, University of Portsmouth, 
University of Utah, University of Virginia, University of Washington, University of Wisconsin, 
Vanderbilt University, and Yale University.

\appendix

{
\section{Characteristic vs. average oxygen abundances}
\label{sec:comp_OH}

In Section \ref{sec:ana} we indicate that the characteristic oxygen
abundance of a galaxy is a good representation of the average oxygen
abundance across the entire optical extension of the galaxy, as
already noticed by \citet{sanchez13}. In order to illustrate it we
show the comparison between these two parameter for the galaxies of
the current analyzed sample for the PP04 calibrator
(Fig. \ref{fig:com_OH_PP04}).  Figure \ref{fig:com_OH} shows a similar
comparison for the remaining ten calibrators discussed along this
article. As expected the characteristic and mean oxygen abundances are
well distributed along a one-to-one relation for the different
calibrators. Even more, it is shown that the typical error for the
mean oyxgen abundance is systematically larger than the one estimated
for the characteristics one. This is expected since the second one is
the result of a linear regression using a large number of individual
values.  In order to quantify how the two parameters compare, we show
in Table \ref{tab:com_OH} the mean and standard deviation of the
differences between them. We include the third decimal, although it is
clearly unsignificant, to highlight any possible difference. However,
it is evident that both parameters are very similar, with a difference
well below the estimated errors. It is important to notice that the
matching between the two estimations is not equally good for the different
calibrators. In general, those calibrators using a larger number of emission
line ratios, or based on high order polynomials of the considered line ratios present
a larger disagreement. However, the calibrator with the larger difference
is that one based on R23, with an offset of $\sim$$-$0.039$\pm$0.051 dex.

\begin{figure*}
 \minipage{\textwidth}
 \includegraphics[clip,trim=100 30 120 100, width=0.33\linewidth]{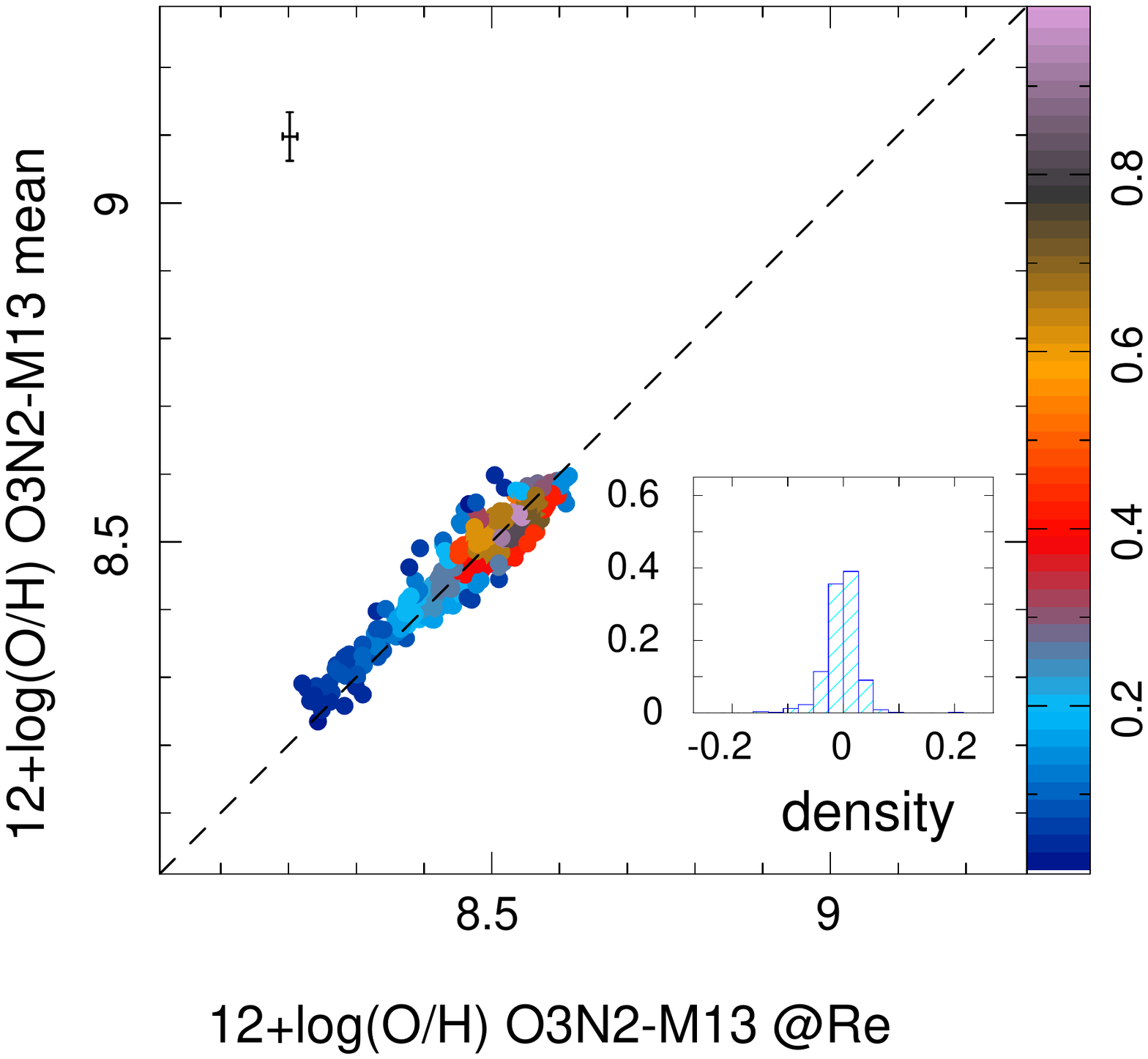}\includegraphics[clip,trim=100 30 120 100, width=0.33\linewidth]{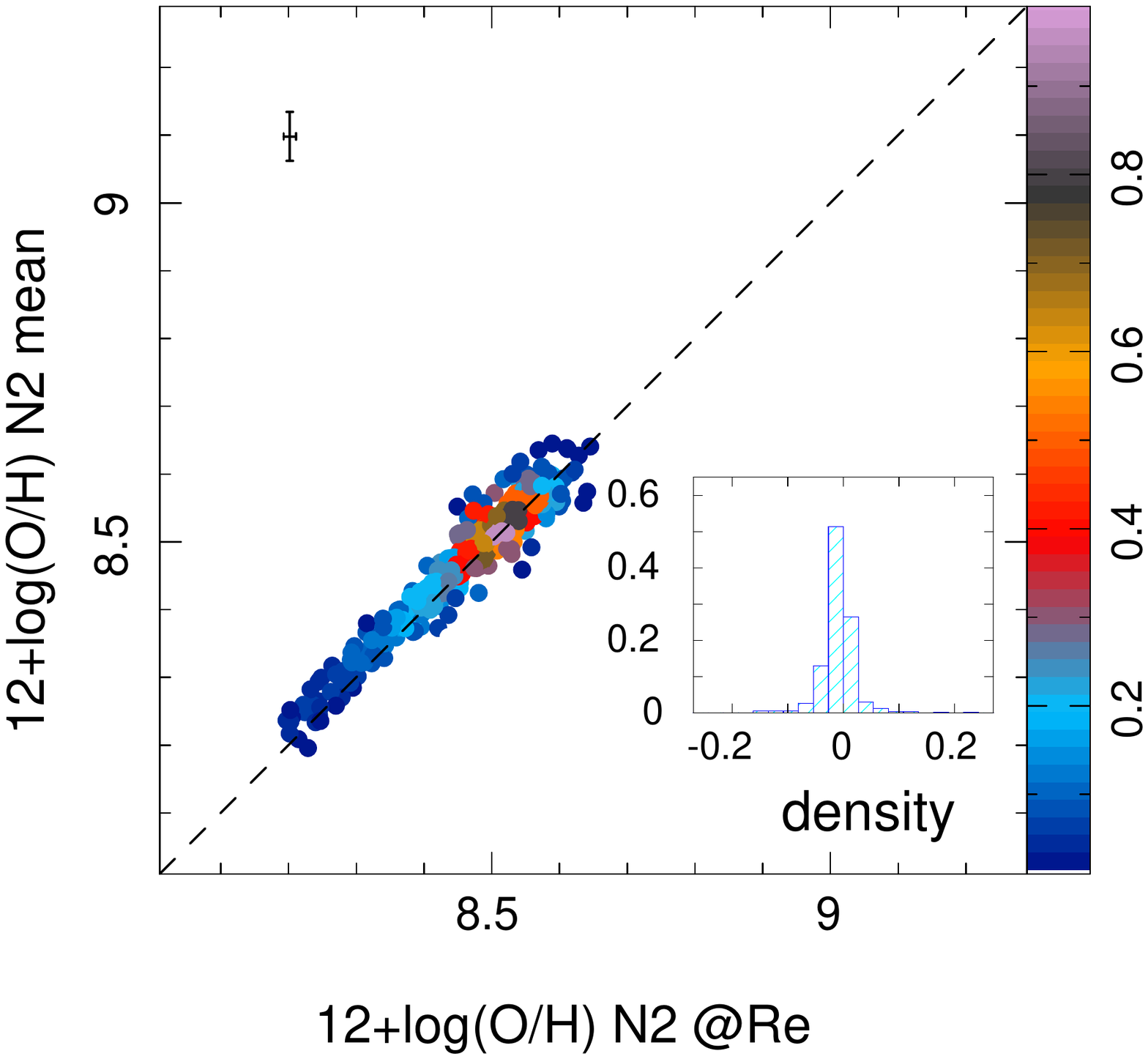}\includegraphics[clip,trim=100 30 120 100, width=0.33\linewidth]{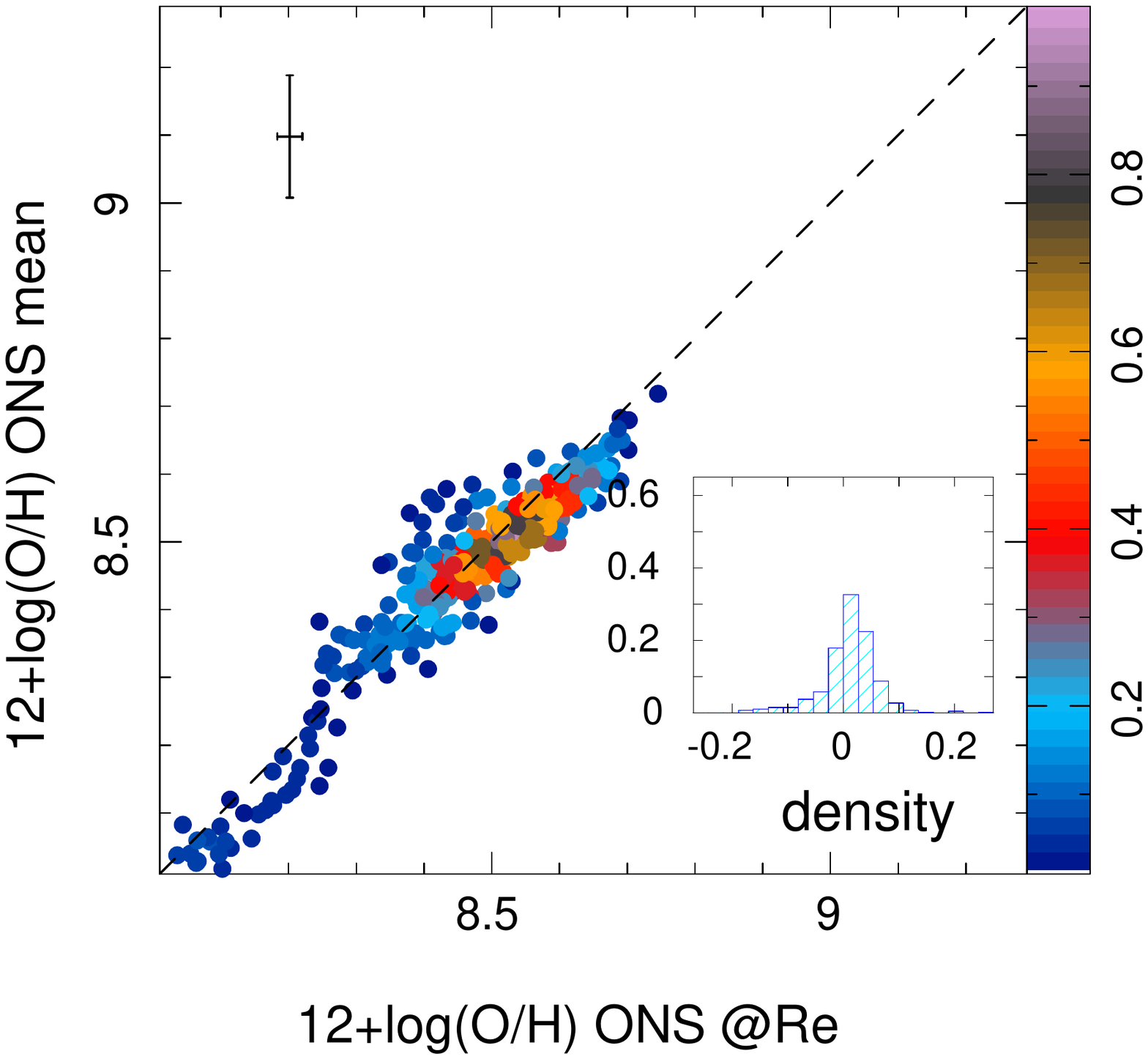}
 \includegraphics[clip,trim=100 30 120 100, width=0.33\linewidth]{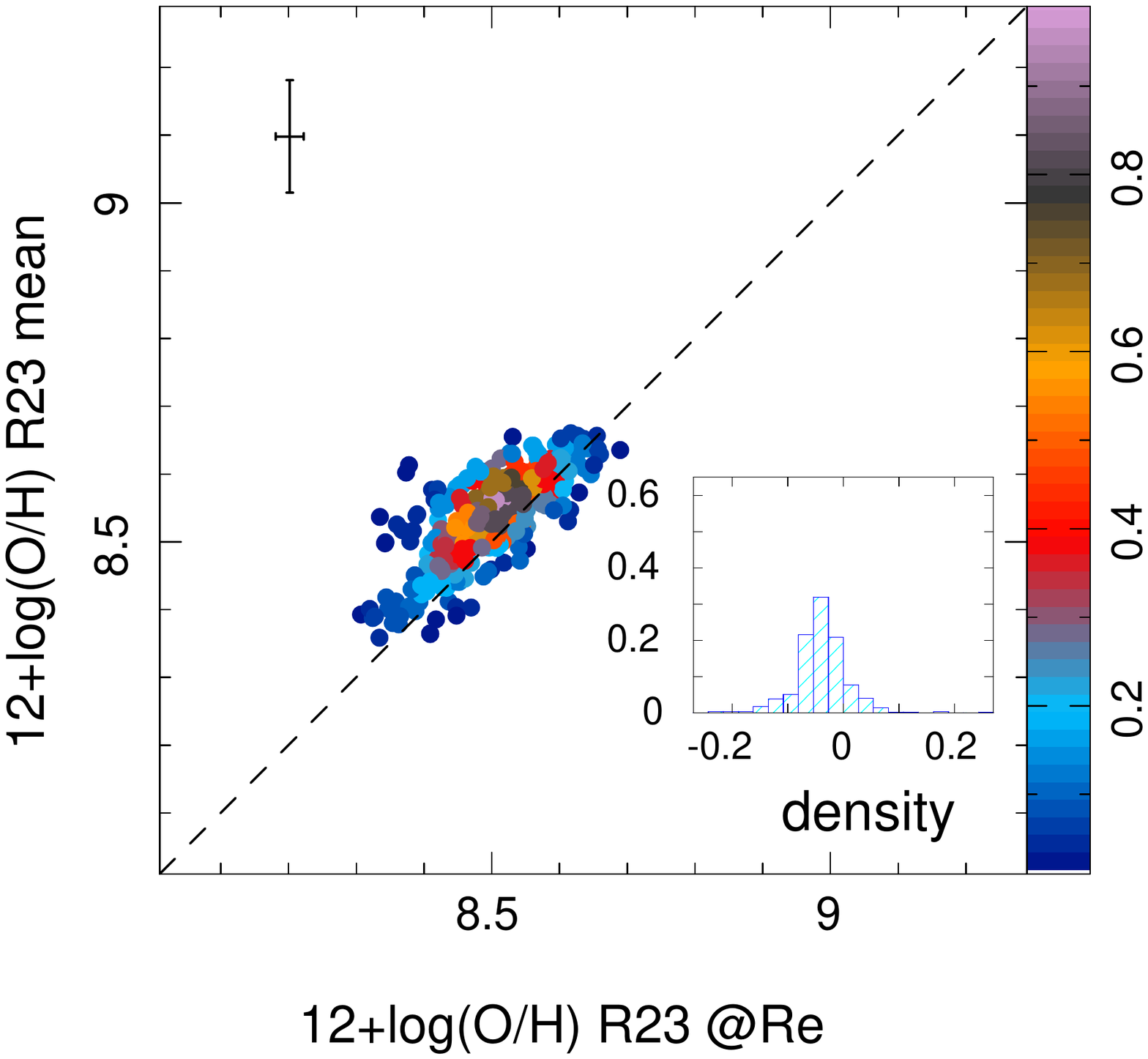}\includegraphics[clip,trim=100 30 120 100, width=0.33\linewidth]{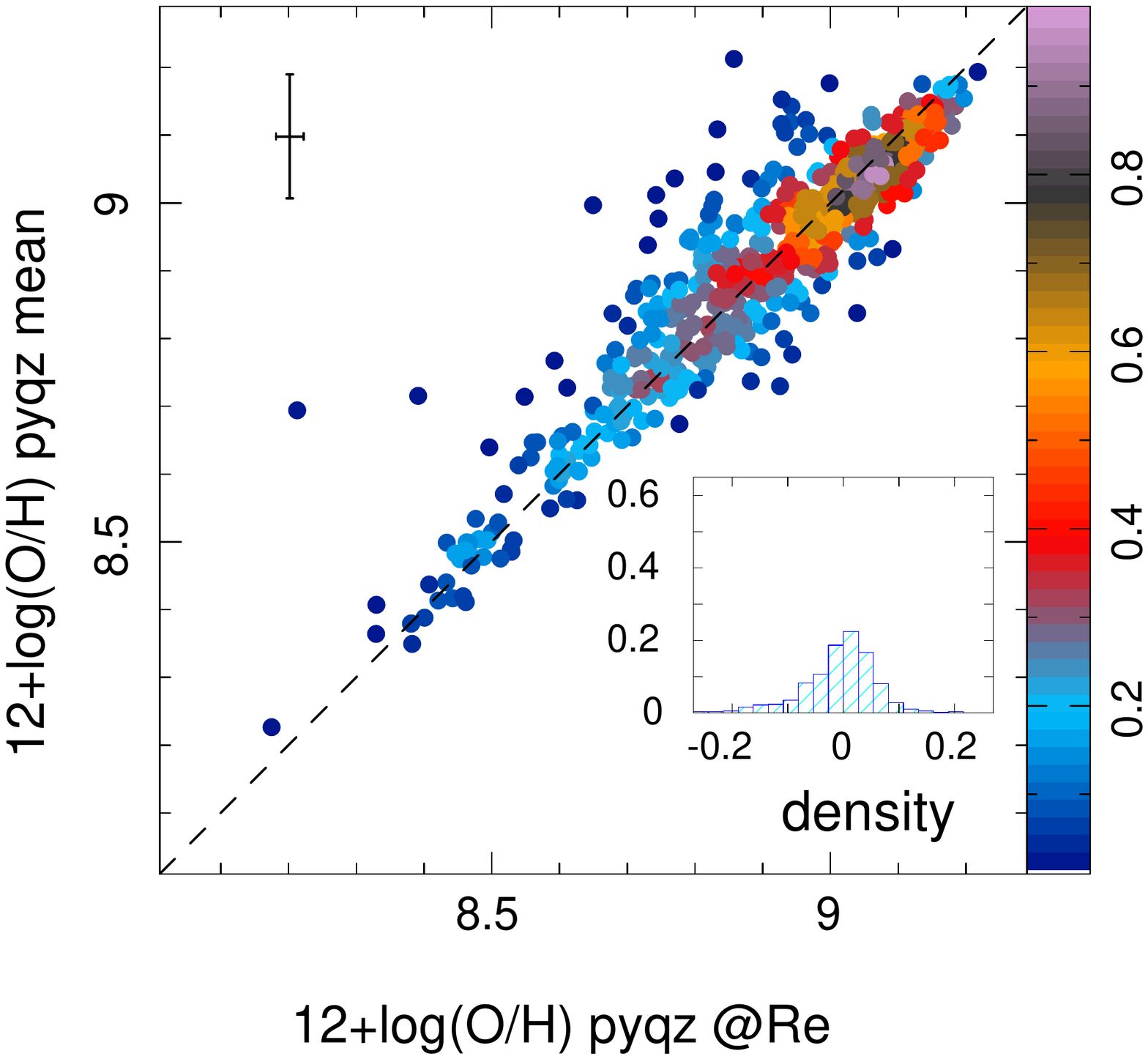}\includegraphics[clip,trim=100 30 120 100, width=0.33\linewidth]{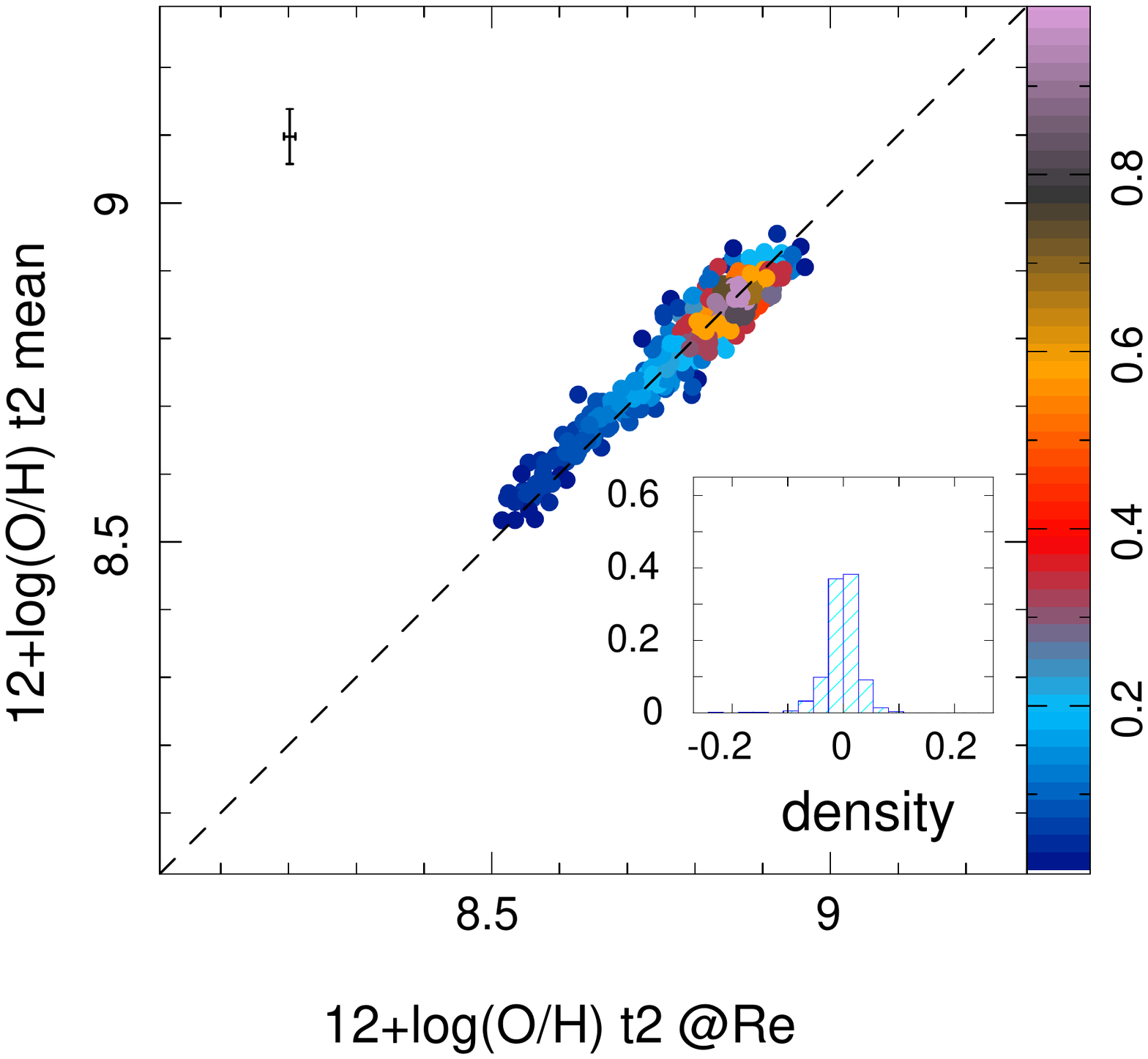}
 \includegraphics[clip,trim=100 30 120 100, width=0.33\linewidth]{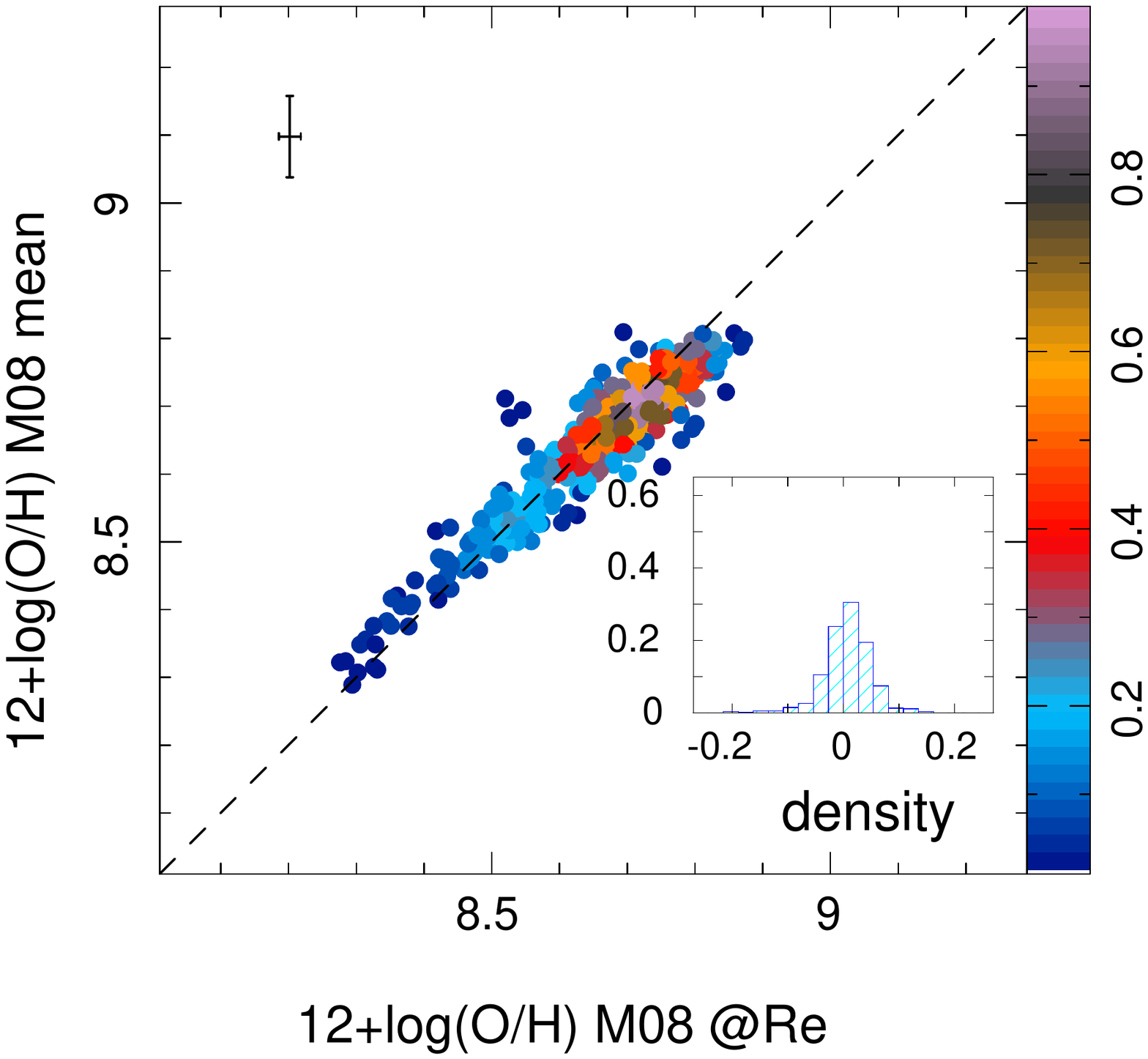}\includegraphics[clip,trim=100 30 120 100, width=0.33\linewidth]{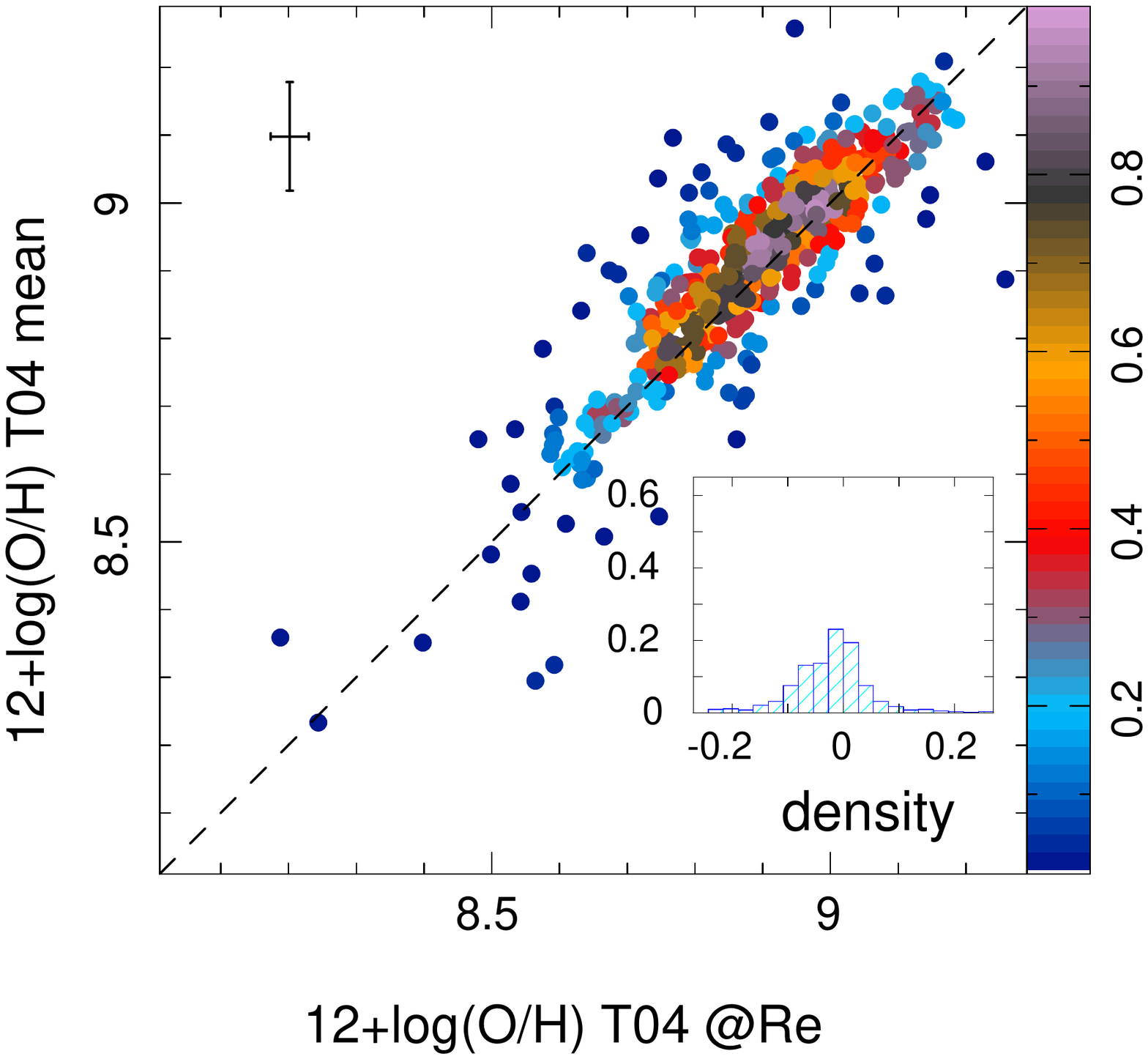}\includegraphics[clip,trim=100 30 120 100, width=0.33\linewidth]{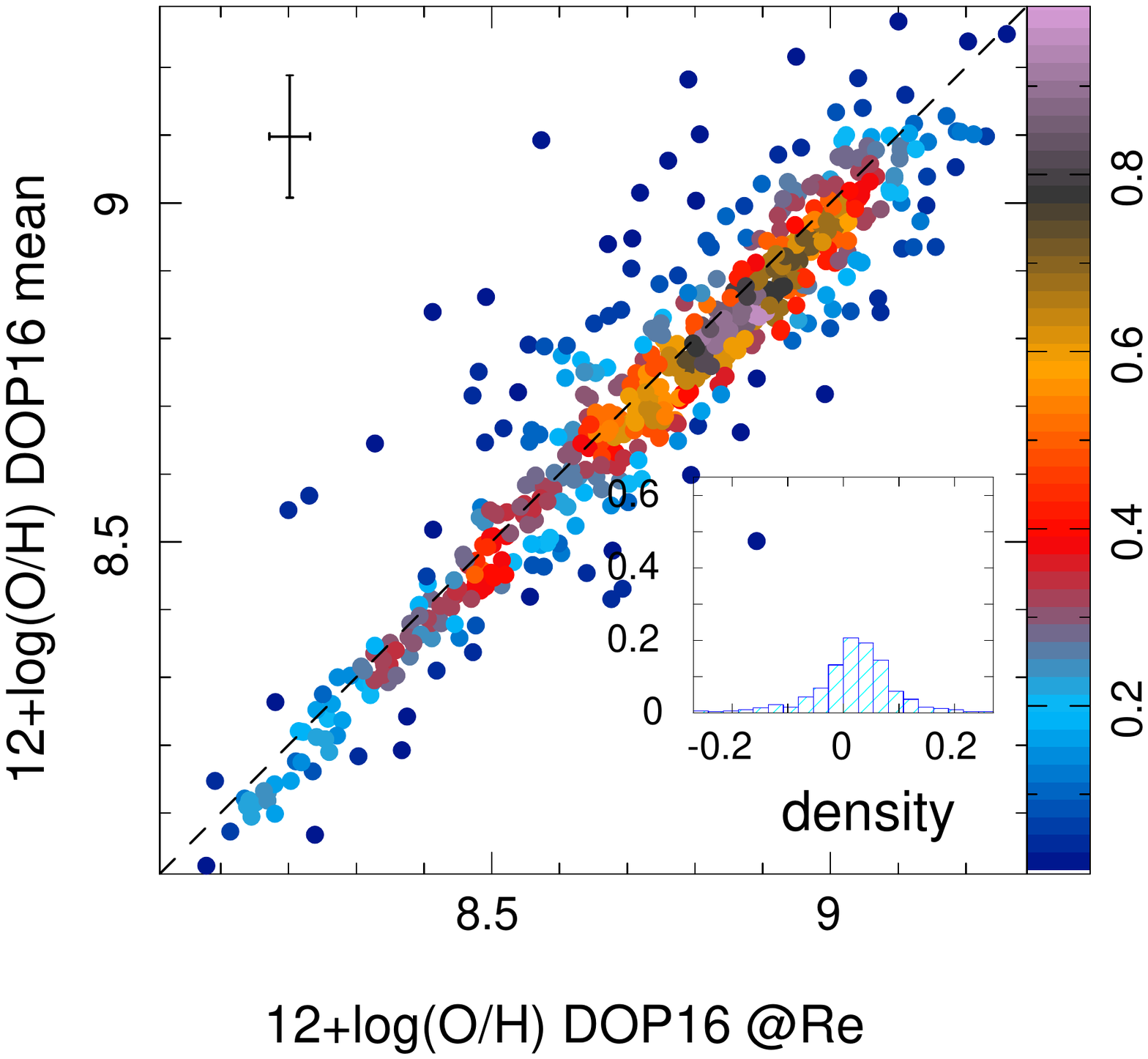}
 \includegraphics[clip,trim=100 30 120 100, width=0.33\linewidth]{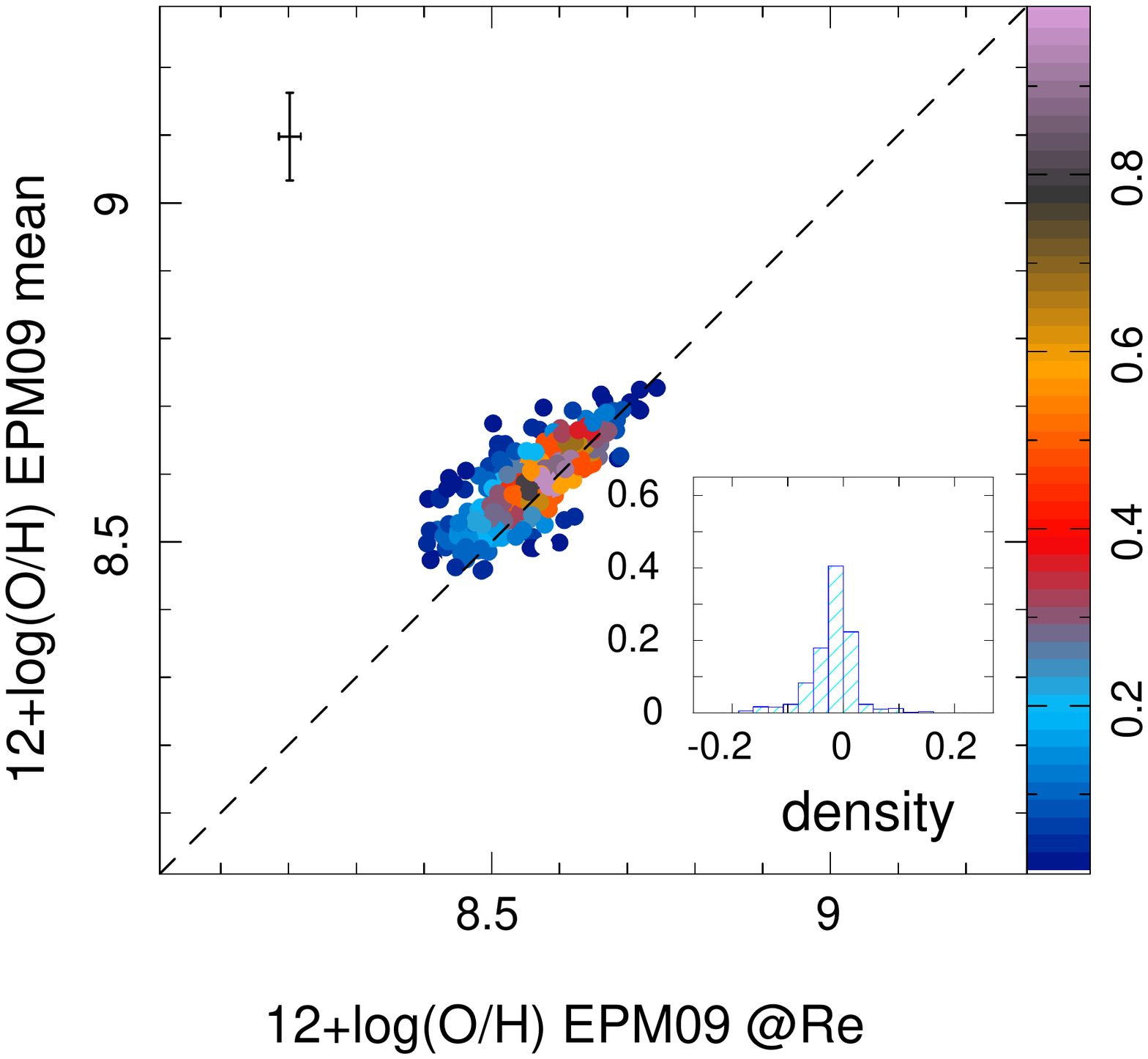}
 \endminipage
\caption{ Comparison between the average oxygen abundance across the entire Field-of-View of the datacubes (beyond $\sim$2.5 r$_e$) and the characteristic oxygen abundance at the effective radius using the linear regression technique described in the text for the different calibrators discussed in this study. Each panel corresponds to a different calibrator. In each panel, solid circles correspond to the values for an individual galaxy, with a color code indicating the density of points normalized to the peak density. The error-bar indicates the median error estimated for each parameter. The inset shows a histogram of the difference between the two estimations of the oxygen abundance.}
 \label{fig:com_OH}
\end{figure*}

\begin{table}
\caption[Comparison between average and characteristic oyxgen abundances.]{Comparison between the average and charactistic oxygen abundances. For each calibrator of the ones discussed in this article we include the mean and standard deviations of the differences between both parameters for the set of galaxies included in our study.}
\label{tab:com_OH}
\begin{tabular} {c r r}
\hline
Metallicity   & $\Delta log(O/H)$& $\sigma_{\Delta log(O/H)}$ \\
 Indicator & (dex) & (dex)  \\
\hline
O3N2-M13 & -0.002 & 0.031\\
PP04     & -0.002 & 0.045\\
N2-M13   & -0.009 & 0.033\\
ONS      &  0.012 & 0.049\\
R23      & -0.039 & 0.051\\
t2       & -0.001 & 0.032\\
M08      &  0.010 & 0.056\\
T04      & -0.017 & 0.116\\
EPM09    & -0.020 & 0.047\\
DOP16    &  0.014 & 0.094\\
\hline
\end{tabular}
\end{table}

}

\section{Dataset}
\label{sec:dataset}

Table \ref{tab:dataset} include the integrated stellar masses and star-formation rates for 734 galaxies included in the analyzed sample, derived as described in Sec. \ref{sec:ana}, together with their nominal errors (i.e., not taken into account systematic errors). For those galaxies for which it was possible to derive the characteristic oxygen abundance using the different calibrators described in Sec. \ref{sec:calibrators} we include each of their corresponding values. When it was not possible to derive a particular oxygen abundance it is marked with a {\sc nan} value. The listed errors for the calibrators are the nominal ones derived from the linear regression described in Sec. \ref{sec:ana}, and for that reason are considerable smaller than the typical errors derived from single aperture spectroscopic studies. The systematic errors associated with each calibrator are not listed, although they were taken into account in the analysis of the data. An electronic version of this table can be downloaded from the CALIFA FTP server: \url{ftp://ftp.caha.es/CALIFA/dataproducts/MZ_CALIFA/published_table.csv}.

%

\begin{table*}
\caption[Stellar Mass, SFR and Oyxgen abundanes]{Stellar Masses, star-formation rates and characteristics abundances for the all the considered calibrators. {\it The full table is available on-line.}}
\label{tab:dataset}
\begin{tabular}{crrrrrrrr}
\hline
CALIFA  & log(M$_*$/M$_\odot$ & log(SFR/M$_\odot$/yr) & \multicolumn{6}{c}{12+log(O/H)}\\ 
\cline{4-9}
NAME    &                     &                       &  O3N2-M13  & N2-M13 & ONS & R23 & O3N2-PP04 & {\sc pyqz} \\ 
        &                     &                       &            & $t2$     & M08    &  T04 & DOP & EPM  \\ 
\cline{4-9}
        2MASXJ01331766+1319567 &  9.38 $\pm$  0.11 & -0.94 $\pm$  0.06 &  8.39 $\pm$  0.02 &  8.36 $\pm$  0.01 &  8.36 $\pm$  0.01 &  8.47 $\pm$  0.03 &  8.56 $\pm$  0.03 &  8.69 $\pm$  0.01\\ 
  &      &      &     &  8.69 $\pm$  0.01 &  8.50 $\pm$  0.02 &  8.86 $\pm$  0.04 &  8.34 $\pm$  0.02 &  8.59 $\pm$  0.04\\ 
\hline
                2MASXJ09065870 & 11.60 $\pm$  0.12 &  0.41 $\pm$  0.06 &  8.47 $\pm$  0.06 &  8.54 $\pm$  0.09 &   nan $\pm$   nan &  8.55 $\pm$  0.50 &  8.68 $\pm$  0.09 &  8.88 $\pm$  0.59\\ 
  &      &      &     &  8.83 $\pm$  0.09 &  8.65 $\pm$  0.30 &  8.99 $\pm$  0.33 &  8.94 $\pm$  1.20 &  8.51 $\pm$  0.46\\ 
\hline
                2MASXJ09591230 & 10.39 $\pm$  0.12 & -1.03 $\pm$  0.07 &   nan $\pm$   nan &   nan $\pm$   nan &   nan $\pm$   nan &   nan $\pm$   nan &   nan $\pm$   nan &   nan $\pm$   nan\\ 
  &      &      &     &   nan $\pm$   nan &   nan $\pm$   nan &   nan $\pm$   nan &   nan $\pm$   nan &   nan $\pm$   nan\\ 
\hline
                2MASXJ12095669 & 10.48 $\pm$  0.12 & -0.84 $\pm$  0.06 &  8.53 $\pm$  0.09 &  8.55 $\pm$  0.07 &   nan $\pm$   nan &  8.58 $\pm$  0.12 &  8.77 $\pm$  0.13 &  8.81 $\pm$  0.20\\ 
  &      &      &     &  8.85 $\pm$  0.07 &  8.67 $\pm$  0.09 &  8.86 $\pm$  0.24 &  8.68 $\pm$  0.38 &  8.66 $\pm$  0.12\\ 
\hline
        2MASXJ15024995+4847010 & 10.43 $\pm$  0.10 & -0.44 $\pm$  0.05 &  8.47 $\pm$  0.02 &  8.46 $\pm$  0.05 &   nan $\pm$   nan &  8.45 $\pm$  0.07 &  8.68 $\pm$  0.03 &  8.68 $\pm$  0.09\\ 
  &      &      &     &  8.80 $\pm$  0.08 &  8.63 $\pm$  0.11 &  8.53 $\pm$  0.38 &  8.68 $\pm$  0.33 &  8.60 $\pm$  0.13\\ 
\hline
        2MASXJ15393305+3205382 & 11.08 $\pm$  0.10 & -0.38 $\pm$  0.06 &  8.51 $\pm$  0.03 &  8.57 $\pm$  0.05 &   nan $\pm$   nan &  8.50 $\pm$  0.11 &  8.73 $\pm$  0.05 &  8.93 $\pm$  0.12\\ 
  &      &      &     &  8.84 $\pm$  0.05 &  8.64 $\pm$  0.06 &  8.88 $\pm$  0.13 &  8.92 $\pm$  0.20 &  8.46 $\pm$  0.16\\ 
\hline
        2MASXJ15570268+3725001 & 10.15 $\pm$  0.10 & -1.35 $\pm$  0.05 &  8.51 $\pm$  0.06 &  8.58 $\pm$  0.09 &   nan $\pm$   nan &  8.58 $\pm$  0.14 &  8.74 $\pm$  0.09 &  8.68 $\pm$  0.25\\ 
  &      &      &     &  8.87 $\pm$  0.03 &  8.68 $\pm$  0.15 &  8.98 $\pm$  0.15 &  8.55 $\pm$  0.56 &  8.66 $\pm$  0.33\\ 
\hline
                2MASXJ16152860 & 10.06 $\pm$  0.08 & -0.61 $\pm$  0.06 &  8.49 $\pm$  0.08 &  8.53 $\pm$  0.14 &   nan $\pm$   nan &  8.51 $\pm$  0.24 &  8.71 $\pm$  0.12 &  8.83 $\pm$  0.23\\ 
  &      &      &     &  8.80 $\pm$  0.04 &  8.69 $\pm$  0.09 &  8.80 $\pm$  0.53 &  8.87 $\pm$  0.67 &  8.58 $\pm$  0.21\\ 
\hline
                        ARP180 & 10.42 $\pm$  0.09 & -0.71 $\pm$  0.07 &  8.46 $\pm$  0.03 &  8.50 $\pm$  0.04 &  8.43 $\pm$  0.07 &  8.51 $\pm$  0.09 &  8.67 $\pm$  0.04 &  8.76 $\pm$  0.12\\ 
  &      &      &     &  8.81 $\pm$  0.04 &  8.64 $\pm$  0.07 &  8.65 $\pm$  0.12 &  8.65 $\pm$  0.12 &  8.58 $\pm$  0.06\\ 
\hline
                        ARP220 & 10.86 $\pm$  0.11 &  1.86 $\pm$  0.06 &  8.56 $\pm$  0.05 &  8.67 $\pm$  0.05 &  8.41 $\pm$  0.07 &  8.74 $\pm$  0.30 &  8.81 $\pm$  0.07 &  9.07 $\pm$  0.11\\ 
  &      &      &     &  8.95 $\pm$  0.03 &  8.74 $\pm$  0.06 &  8.64 $\pm$  0.28 &  9.07 $\pm$  0.18 &  8.57 $\pm$  0.11\\ 
\hline
\end{tabular}
\end{table*}

\bibliographystyle{mnras}
\bibliography{main,CALIFAI}

\end{document}